%% file: main.tex
\pdfoutput=1
\documentclass[sigconf,screen]{acmart}

\AtBeginDocument{%
  }

\renewcommand\footnotetextcopyrightpermission[1]{} 
\usepackage{xspace}

\newcommand{\magic}{\textsc{SplitBFT}\xspace} 

\newcommand{\showExtra}[1]{}
\usepackage{stfloats}

\usepackage{acronym}
\usepackage{todonotes}
\usepackage{tabularx}
\usepackage{booktabs}
\usepackage{cleveref}

\usepackage{pifont}

\newcommand{\cmark}{\textcolor{green}{\ding{51}}}%
\newcommand{\xmark}{\textcolor{red}{\ding{55}}}%

\usepackage{spacetime}

\acrodef{SGX}{Software Guard Extensions}
\acrodef{SDK}{Software Development Kit}
\acrodef{PRM}{Processor Reserved Memory}
\acrodef{EPC}{Enclave Page Cache}
\acrodef{EDL}{Enclave Description Language}
\acrodef{TEE}{trusted execution environment}
\acrodef{RA}{Remote Attestation}
\acrodef{TCS}{Thread Control Structure}
\acrodef{SSA}{State Save Area}
\acrodef{AEX}{Asynchrounous Enclave Exit}
\acrodef{AEP}{Asynchronous Exit Pointer}
\acrodef{IDEA}{International Data Encryption Algorithm}
\acrodef{BaaS}{Blockchain-as-a-Service}
\acrodef{BFT}{Byzantine Fault Tolerance}
\acrodef{PBFT}{Practical Byzantine Fault Tolerance}
\acrodef{TEE}{Trusted Execution Environments}
\acrodef{EPC}{Enclave Page Cache}
\acrodef{YCSB}{Yahoo Cloud Serving Benchmark}
\usepackage[plain]{algorithm}
\usepackage{algpseudocode}
\usepackage{stackengine}
\usepackage{tikz,calc}
\usepackage{cancel}
\usepackage{graphicx}
\usetikzlibrary{tikzmark,calc}

\usepackage{multirow}
\usepackage{xcolor}

\usepackage{multirow}

\usepackage{hyperref}
\usepackage{tikz}
\usepackage{xcolor}
\usepackage{listings}
\usepackage[frozencache,cachedir=.]{minted}
\usepackage{subfigure}

\definecolor{kaiYellow}{RGB}{204,204,0}

\definecolor{aliceblue}{rgb}{0.94, 0.97, 1.0}
\definecolor{mintgreen}{rgb}{0.6, 1.0, 0.6}

\usepackage{tikz}
\newcommand*\circl[1]{\tikz[baseline=(char.base)]{
            \node[shape=circle,draw,inner sep=1pt, fill=aliceblue] (char) {#1};}}

\newcommand*\cir[1]{\tikz[baseline=(char.base)]{
            \node[shape=circle,draw,inner sep=1pt, fill=white, text=black] (char) {#1};}}
\newcommand{\lib}[1]{\texttt{#1}}

\definecolor{DarkGreen}{rgb}{0,0.43,0}
\definecolor{mGreen}{rgb}{0,0.7,0}
\definecolor{moeColor}{rgb}{168,0,60}
\definecolor{mGray}{rgb}{0.5,0.5,0.5}
\definecolor{mPurple}{rgb}{0.58,0,0.82}
\definecolor{mBlue}{RGB}{0,0,255}
\definecolor{backgroundColour}{rgb}{0.97,0.97,0.97}

\lstdefinestyle{C++Style}{
	language=C++,
	backgroundcolor=\color{backgroundColour},
	commentstyle=\color{mGreen},
	keywordstyle=\color{mGreen},
	numberstyle=\tiny\color{mGray},
	stringstyle=\color{mPurple},
	basicstyle=\footnotesize,
	breakatwhitespace=false,
	breaklines=true,
	captionpos=b,
	keepspaces=true,
	numbers=left,
	numbersep=5pt,
	showspaces=false,
	showstringspaces=false,
	showtabs=false,
	tabsize=2,
	classoffset=1, 
	otherkeywords={user\_check, string, in\ , out, uint64\_t, from, import},
	morekeywords={user\_check,  string, in\ ,out, uint64\_t, from, import},
	keywordstyle=\color{mBlue},
	classoffset=0
}
\usepackage[normalem]{ulem}

\algnewcommand\algorithmicforeach{\textbf{for each}}
\algdef{S}[FOR]{ForEach}[1]{\algorithmicforeach\ #1\ \algorithmicdo}

\algdef{SE}[SUBALG]{Indent}{EndIndent}{}{\algorithmicend\ }%
\algtext*{Indent}
\algtext*{EndIndent}

\newcommand{\princ}[1]{\circl{$P_{#1}$}}

\newcommand{\prepare}{\textsc{prepare}}

\newcommand{\commit}{\textsc{commit}}

\newcommand{\preparationC}{\textit{Preparation}\xspace}
\newcommand{\executionC}{\textit{Execution}\xspace}
\newcommand{\confirmationC}{\textit{Confirmation}\xspace}

\newcommand{\prepreparemsg}{\textsc{PrePrepare}\xspace}
\newcommand{\preparemsg}{\textsc{Prepare}\xspace}
\newcommand{\commitmsg}{\textsc{Commit}\xspace}

\newcommand{\checkpointmsg}{\textsc{Checkpoint}\xspace}
\newcommand{\viewchangemsg}{\textsc{ViewChange}\xspace}
\newcommand{\newviewmsg}{\textsc{NewView}\xspace}

\newlength\myindent
\algdef{SE}[SUBALG]{Indent}{EndIndent}{}{\algorithmicend\ }%
\algtext*{Indent}
\algtext*{EndIndent}

\newcommand{\mypar}[1]{\paragraph{#1}}
\newcommand{\myparinline}[1]{\textbf{#1.}\xspace}

\newenvironment{myitemize}{%
\begin{itemize}[leftmargin=1em, itemsep=.1em, parsep=.1em, topsep=.1em,
    partopsep=.1em]}
{\end{itemize}}

\newenvironment{myenumerate}{%
\begin{enumerate}[leftmargin=1.50em, itemsep=0em, parsep=0em, topsep=.1em,
    partopsep=.1em]}
{\end{enumerate}}

\newenvironment{structure*}{\color{blue}\begin{myenumerate}}{\end{myenumerate}}
\begin{document}

\title{\magic: Improving Byzantine Fault Tolerance Safety Using Trusted Compartments}

\author{Ines Messadi}
\affiliation{%
	\city{TU Braunschweig}
	\country{Germany}
}

\author{Markus Horst Becker}
\affiliation{%
	\city{TU Braunschweig}
	\country{Germany}
}

\author{Kai Bleeke }
\affiliation{%
	\city{TU Braunschweig}
	\country{Germany}
}

\author{Leander Jehl}
\affiliation{%
	\city{University of Stavanger}
	\country{Norway }
}
\authornote{Work done while at TU Braunschweig - Germany}

\author{Sonia Ben Mokhtar}
\affiliation{%
	\city{LIRIS-CNRS}
	\country{France}
}

\author{R\"udiger Kapitza}
\affiliation{%
	\city{TU Braunschweig}
	\country{Germany}
}

\renewcommand{\shortauthors}{Messadi et al.}


\begin{abstract}
	\input{sections/00-abstract.tex}

\end{abstract}






\maketitle

\input{sections/01-intro}

\input{sections/02-background}

\input{sections/03-design}

\input{sections/04-implemtation}
\input{sections/05-evaluation}
\input{sections/06-relatedwork}

\input{sections/07-conclusion}

\section*{acknowledgment}
We thank our anonymous reviewers for their helpful comments. This work was supported
by the German Research Foundation (DFG) under grant no. KA 3171/9-1.

\bibliographystyle{ACM-Reference-Format}
\bibliography{main.bib}

\end{document}

%% file: sections/00-abstract.tex


Byzantine fault-tolerant agreement (BFT) in a partially synchronous system usually requires $3f+1$ nodes to tolerate $f$ faulty replicas. 
Due to their high throughput and finality property BFT algorithms build the core of recent permissioned blockchains.
As a complex and resource-demanding infrastructure, multiple cloud providers have
started offering  \acl{BaaS}.
This eases the deployment of permissioned blockchains but places the cloud
provider in a central controlling position, thereby questioning blockchains' fault tolerance and decentralization properties and their underlying BFT algorithm.

This paper presents \magic, a new way to utilize trusted execution technology (TEEs), such as Intel SGX, to harden the
safety and confidentiality guarantees of BFT systems thereby strengthening the trust in could-based deployments of permissioned blockchains.   
Deviating from standard assumptions, \magic{} acknowledges that code protected by trusted execution may fail.
We address this by splitting and isolating the core logic of BFT protocols into multiple compartments resulting in a more resilient architecture.
We apply \magic to the traditional practical byzantine fault tolerance algorithm (PBFT) and evaluate it using SGX.
Our results show that \magic adds only a reasonable overhead compared to the non-compartmentalized variant.


%% file: sections/01-intro.tex
\section{Introduction}
\ac{BFT} agreement algorithms are designed to tolerate arbitrary failures in distributed systems~\cite{castro1999practical,lamport2019byzantine}.
Over the years, they have been extended to become faster~\cite{dumbo, sbft}, flexible~\cite{bahsoun2015making,malkhi2019flexible}, resource-efficient~\cite{cheapbft,behl2017hybrids}, and in many more ways. 
Recently, they started receiving increasing attention as they order transactions at the heart of many blockchain infrastructures~\cite{androulaki2018hyperledger,libra}. 

Blockchains were initially focused on exchanging digital currencies in a trustless, decentralized manner. 
Yet, their purpose broadened to enable secure transactions of all kinds, e.g., supply chains, NFT marketplaces, secure sharing of medical data~\cite{retail, kuo2017blockchain, usecases}. 
Because setting up a blockchain infrastructure for such purposes is a non-trivial effort, several companies (e.g., cloud providers such as Microsoft Azure~\cite{azure} and Amazon AWS~\cite{amazonblockchain}) launched \ac{BaaS} to spread their blockchain solution while providing the underlying infrastructure. 
Specifically, they offer public open (also called \emph{permissionless}) blockchains for use cases in which users can participate unrestrictedly and \emph{permissioned} blockchains for more sensitive use cases that require a restricted access control list, that allows only authorized participants.


%
However, having a central blockchain service provider contradicts one of the pillars that made the success of block\-chains, i.e., decentralized governance, and it partly jeopardizes the fault model of the utilized BFT algorithm.
Indeed, the \ac{BaaS} provider does not only become a single point of failure, possibly harming the blockchain service availability, but it also requires strong trust in the provider and his personnel regarding the integrity and confidentiality of the hosted blockchain. 
  

For conventional applications, recent \ac{TEE} remedy the trust issues in terms of integrity and confidentiality of the cloud, as these environments provide an isolated and shielded execution, protecting applications even from local privileged attackers~\cite{costan2016intel}.
Today, TEEs (e.g., Intel Software Guard Extension) are readily available in cloud infrastructures, and several applications have been extended for its use~\cite{confidentialc}. 

 TEEs have already been identified as a technical means to improve the performance and resource-efficiency of \ac{BFT} in conventional data center settings~\cite{behl2017hybrids,minbft,cheapbft}. 
However, for these systems, the main idea is to address \emph{equivocation} based on a \emph{hybrid fault model}~\cite{cheapbft, minbft, behl2017hybrids}, where malicious replicas send conflicting messages during agreement. 
In these models, a fraction of the codebase is shielded by trusted execution that is assumed to fail only by crashing and by definition, cannot be subject to Byzantine faults. 
While introducing trusted execution for BFT, such as in the hybrid fault model, can be beneficial, it does not address the integrity and confidentiality issues of current \ac{BaaS}-based solutions where the cloud provider has the central control over the infrastructure.

This paper introduces \magic, a TEE-tailored BFT architecture enabling a more trustworthy implementation of \ac{BaaS}.
Specifically, \magic introduces an agreement protocol's fine-grained, TEE-based compartmentalization.
Splitting a BFT protocol into TEE-enabled compartments improves resilience and confidentiality and eases the implementation of diversity. 

We illustrate \magic by compartmentalizing \ac{PBFT}. Using our solution, PBFT is decomposed into three independent compartments guarded by trusted execution. 
The separation follows a careful analysis enforcing that individual steps in the protocol are secured by a quorum decision that makes these steps independent of each other.
Based on this approach, the resilience of \magic goes beyond traditional BFT protocols in terms of integrity by tolerating $f$ Byzantine faults of a particular compartment type. 
In terms of confidentiality, the service tolerates Byzantine faults as long as a specific type of compartment that hosts the service is correct. 
Also  \magic contributes to easing the diversity as it limits the code base that has to be diversified at the level of the BFT protocol implementation to preserve integrity. 
Orthogonal to these improvements, \magic ensures availability if not more than $f$ nodes are faulty.
 
\par{We implemented \magic as a Rust-based framework that offers a compartmentalized version of PBFT and utilizes Intel \ac{SGX} as a trusted execution technology.
However, \magic is a generic approach to compartmentalizing BFT protocols that is neither limited to PBFT nor \ac{SGX}.
We evaluate \magic in a cloud setting with two use cases: (i) the replication of a trusted key/value store and (ii) as an ordering service for a blockchain application. 
Both use cases show moderate overhead compared to a plain use of PBFT, which instead offers weaker integrity, no confidentiality protection, and is harder to diversify.

\noindent In summary, this paper makes the following contributions: 

\begin{itemize}
\item \myparinline{Compartmentalized BFT} We present the first approach towards a multi-compartment paradigm for BFT applications using SGX enclaves and show how a \textit{compartmentalized} system that uses multiple enclaves can improve safety and preserve integrity with more than $f$ adversaries. We also propose principles that similar BFT protocols can apply.

\item \myparinline{Diversity of Replicas} We ease the diversity using multiple compartments that share no common code to minimize the number of shared vulnerabilities among enclaves. Furthermore, in our model, enclaves can fail independently.

\item \myparinline{Cloud-tailored BFT} We propose a confidentiality-enhanced BFT protocol tailored for consortium block\-chain in a cloud setting. We present an extensive experimental evaluation of our system in scenarios with up to 150 clients using two applications; a key-value store and a distributed ledger.

\end{itemize}

%% file: sections/02-background.tex
\section{Towards a Fine-Grained Partitioning of BFT Protocols}\label{sec:background}

\mypar{The crux of BFT Protocols} 
\ac{BFT} spanned years of academic research but noticeably gained renewed interest due to its relevance in permissioned blockchains where nodes must be authorized~\cite{yin2019hotstuff,androulaki2018hyperledger}, including more recent cases of \acl{BaaS}~\cite{azure,amazonblockchain} where a cloud provider hosts the entire blockchain infrastructure for its customers. In a nutshell, a BFT consensus typically runs between nodes assuming a broad spectrum of faults, covering arbitrary and malicious behavior.
Thus, they are employed to allow cooperation between participants that do not trust each other or in scenarios where individual replicas may be subject to unpredicted faults or outside attacks. The fault model of BFT systems is extensive and comes at a high cost: typically, at most $f< n/3$ out of $n$ replicas can be faulty.
This differs from systems focused on crash failures, where $f < n/2$ can be tolerated.
Accordingly, when deploying a BFT system, it is essential to consider failure assumptions.
Especially, common failures that may happen on all replicas must be prevented.
We argue that for cloud applications, like \ac{BaaS}, \ac{BFT}'s basic fault model is hard to maintain, considering the higher likelihood of an attacker or a rogue administrator that controls more than a third of the nodes. Furthermore, traditional BFT designs assume no confidentiality, an essential ingredient for cloud users and sensitive blockchain applications.

\input{fig/failureModelTable.tex}

\mypar{TEE in BFT} Security threats in cloud infrastructure motivated the use of CPU extensions that support trusted execution (\acp{TEE}) in various applications, including BFT~\cite{kim2019shieldstore,arnautov2016scone}.
BFT protocols leverage Intel SGX as it allows trusted applications with small TCB and for its maturity and availability in commodity hardware compared to its predecessor TEEs~\cite{behl2017hybrids,minbft}.
Intel SGX's isolated execution environments are known as \emph{enclaves}, a hardware-protected and encrypted memory. It offers strong integrity and confidentiality protection from the underlying OS. A fundamental property of SGX is creating multiple separate enclaves on one CPU. 
Thus, it gives three opportunities for BFT protocols:

\begin{itemize}
    \item[\cir{$o_{1}$}] \textit{Integrity-protection.} it allows to guard the integrity of state and code loaded into the enclave from faults and adversaries in the environment of the replica;
    \item[\cir{$o_{2}$}] \textit{Inter-enclave protection.} it safeguards code in one enclave from failures in a different enclave and;\par
    \item[\cir{$o_{3}$}]  \textit{Confidentiality.} it protects the confidentiality of the data in the enclave from failures and attackers outside that enclave. 
\end{itemize}

\noindent

Clement et al.~\cite{clement2012limited} have shown that, relying on \cir{$o_{1}$}, $f$ replica failures can be tolerated using only 
$2f+1$ replicas. They require the use of digital signatures and a TEE that may only fail by crashing.
So-called \emph{hybrid} systems~\cite{behl2017hybrids,cheapbft} 
use a minimal trusted subsystem, e.g., a trusted counter, to implement this design. However, like traditional BFT protocols, hybrid systems lose safety and liveness if an attacker gains access to more than $f$ replicas or if an enclave within any replica acts maliciously or deviates from the protocol. 
Specifically, while secure, SGX enclaves are susceptible to attacks, significantly when increasing the TCB. 
The application code itself may include bugs and memory corruption that leads to vulnerabilities and security vulnerabilities~(e.g., synchronization bugs~\cite{weichbrodt2016asyncshock}). Most applications that rely on SGX use the memory-unsafe C/C++ because of the C-based interfaces of the SGX SDK. In this matter, the SGX hardware gives no memory-safety guarantees for the software running in the enclave. Using Rust as a memory-safe alternative improves application code safety, but it is not entirely safe from memory corruption errors~\cite{teerex}.
In hybrid approaches, a single byzantine fault, e.g., a bug or successful attack breaching the trusted subsystem, puts safety at risk.

\mypar{\magic}
Unlike hybrid protocols, we do not aim to use TEEs to reduce BFT protocols' overhead but rather to increase BFT resilience by enforcing safety despite more complex failure scenarios. 
We exploit Opportunity \cir{$o_{1}$} to guarantee safety despite an attacker being present on all machines. To achieve this, we place both the core BFT logic and the execution of client requests into the trusted environment.

\begin{figure}[t]

    \flushleft\includegraphics[width=0.45\textwidth]{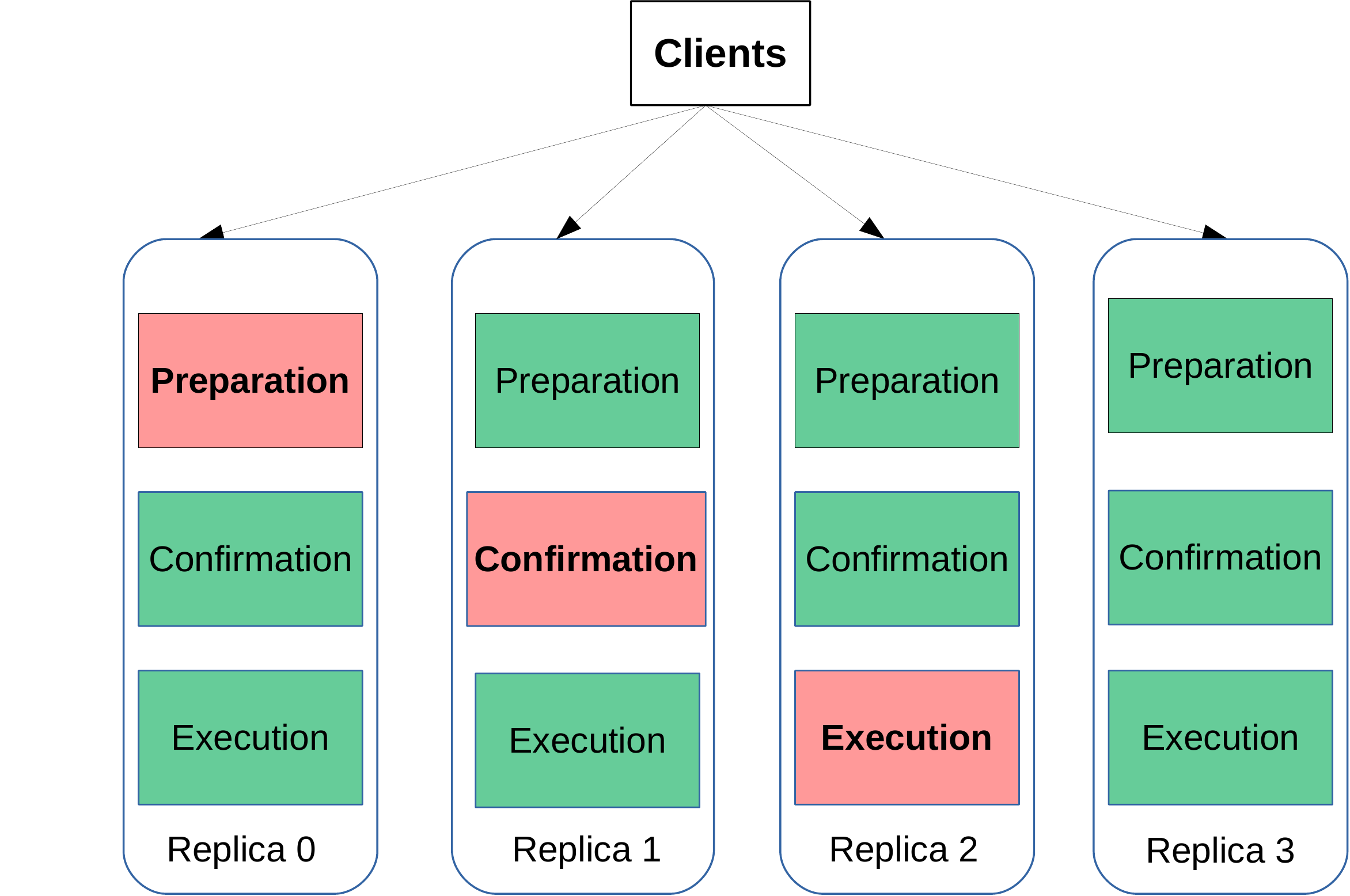}
    
    \caption{Compartmentalized PBFT with four nodes.  Using \magic, failures indicated in red, in different compartments and on multiple replicas can be tolerated.}
    \label{figure:components}
\end{figure}

Further, we do assume that enclaves can fail and become byzantine, i.e., enclaves can equivocate. \magic design tolerates any $f$ enclaves failing. 
To further increase reliability, we aim for a small TCB. Especially, using Opportunity \cir{$o_{2}$}, we separate the complex protocol logic located at each replica into multiple independent compartments, each run in a different enclave.
This allows us to even tolerate faults in more than $f$ individual enclaves, given that these happen in enclaves of different compartment types.
Figure~\ref{figure:components} shows our compartmentalization of the PBFT algorithm. With four replicas, safety is ensured, even if one enclave of each type is faulty.

Finally, we encrypt client requests and responses and only decrypt them inside the enclave.
Thus, we utilize Opportunity \cir{$o_{3}$} to ensure confidentiality despite an attacker present on one of the replicas, as long as enclaves are fault-free. 
In our PBFT variant from Figure~\ref{figure:components}, confidentiality is maintained as long as all enclaves of type \executionC are correct.

In Table \ref{tab:my-table}, we compare \magic to previous plain BFT protocols such as \ac{PBFT} and prevailing TEE-based BFT protocols, i.e., hybrid protocols. \magic follows traditional BFT protocols liveness-wise; however, it enhances resilience. Our design guarantees safety despite $n$ powerful attackers. In other words, an attacker on each replica host, which is not the case for hybrid protocols and PBFT that can only tolerate up to $f$ byzantine attacks. 
\magic also tolerates faulty enclaves. Enclaves belong to different compartment types, and $f$ enclaves belonging to each compartment type may fail. 
In summary, \magic guarantees safety in cases where PBFT and hybrid protocols do not. 
\magic separates liveness from safety and tries to ensure safety even when liveness may be lost. Violating safety provides the attacker gains, such as double-spending attacks or clients receiving inconsistent replies but violating liveness does not. Losing liveness means progress happens only at the disposal of Byzantine replicas but may still happen or be regained later.

Finally, we target confidentiality of client requests, a property that is important in a cloud setting.
In \magic, the client's requests and the application state remain confidential despite an attacker corrupting the environment on one or multiple machines.

\mypar{Diversification}
As mentioned, traditional BFT systems need to avoid common failures of multiple replicas.
Typically, this is achieved by diversifying the replicas' environment.
Diversity is not trivial due to the complexity of implementations and the wide range of employed techniques and libraries. The most common approach is to use different and diverse OSs~\cite{garcia2019lazarus}. 
However, diversifying the implementation remains a tedious and error-prone task, especially when unsafe languages such as C are susceptible to memory errors, e.g., buffer overflows. 
\magic's fault model enables a more accessible approach that diversifies only a fraction of the codebase, namely the code running inside the different enclaves. The use of compartments further simplifies diversification; only common failures in the same compartment need to be avoided.

\subsection{System Assumptions}

We use a similar networking model as presented in the traditional \ac{BFT} state machine replication model~\cite{yin2003separating, castro1999practical}. However, the goal is to tolerate many simultaneous replica faults under pre-defined timing and threshold assumptions.
The network is unreliable and may discard, reorder, and delay messages but not indefinitely to guarantee liveness.
Safety guarantees do not depend on timing or crash failures.

\par
\textit{Modeling \magic.}  
The system consists of a set of clients $\mathcal{C}$ and a set of replicas $\Pi = \{ r_1, r2, ..., r_n \}$. 
We deviate from classical BFT assumptions by not treating a replica as a unit of failure.
Instead, we assume that a replica $r_i$ contains multiple compartments (enclaves) $C = {\{c_{i}^1,c^2_{i},...\}}$ and an environment $e_i$ such as $r_i= C \cup {e_i}$.
Both the environment and compartments may fail arbitrarily. Thus, a faulty or \emph{byzantine compartment} may omit or delay operations, perform arbitrary steps and lie to other compartments.
We note that a fault in the environment may render the compartments in that replica unavailable.
The environment may fail independently of the compartments, while compartments, whether part of the same or different replicas, fail independently. However, we assume that once a compartment (e.g., $c_i^1$) is faulty, the environment $u_i$ is considered to be faulty (e.g., faulty compartments corrupt the environment)

We consider multiple compartment types, and every replica contains one compartment of each type. Compartments of the same type contain the same logic. Figure~\ref{figure:components} shows the 3 types in our adaptation of PBFT, namely \preparationC, \confirmationC, and \executionC compartments.
In the remainder of this paper, we use the term enclave to denote a compartment part of a specific replica. On the other hand, we use the term compartment to indicate the code or logic executed by all enclaves of a given compartment type.


An attacker may simultaneously compromise the $n$ machines, manipulating separate compartments but not multiple enclaves of the same type. Still, the risk of compromising similar compartments is somewhat reduced by using diverse enclaves environments (e.g., various programming languages), therefore reducing common vulnerabilities. 
We assume Intel SGX enclaves implementation and standard cryptographic operations are correct. Still, the developer code may include programming errors/bugs (e.g., buffer overflow) that can lead to arbitrary enclave behavior if exploited. 
We exclude side-channels on \ac{TEE}s~\cite{274699,263816,shih2017t,weichbrodt2016asyncshock, van2018foreshadow, kocher2019spectre} from our threat model, but mitigations or microcode updates can be applied to \magic ~\cite{oleksenko2018varys,seo2017sgx}. 
We assume that each enclave has a public and private key pair and that private keys of correct enclaves cannot be derived by either the environment or other enclaves on the same replica. In addition,
clients are authorized and identified through public keys.

%% file: fig/failureModelTable.tex
\begin{table*}[t]

    \begin{tabular}{@{}llcccc|cc|c@{}}
        \toprule
        
        \multicolumn{1}{c}{\multirow{2}*{\textbf{Work}}}                            & \multirow{2}*{\textbf{\# Replicas}}  & \multirow{2}*{\textbf{TEE}} & \textbf{Faulty}   & \multirow{2}*{\textbf{Liveness}} & \multicolumn{2}{c}{\hspace{12mm}\textbf{Integrity}} & \multicolumn{2}{c}{\textbf{Confidentiality}} \\ 
                                                                        &                                      &                             & \textbf{TEE}      &                                  & \textbf{Enclave} & \textbf{Host}     &  \textbf{Enclave} & \textbf{Host}\\             
        \hline
        PBFT~\cite{castro1999practical}                                 &  \hspace{4mm}$3f+1$   & \xmark              & -                        & $f$                                          & -                                          &  $f$                        &    - &    0            \\
        Hybrid Protocols~\cite{behl2017hybrids,cheapbft}     & \hspace{3mm} $2f+1$   & \cmark              & \xmark                   & $f$                                          & 0                                          &  $f$                        &    0 &    0           \\ \bottomrule
        \magic                                                          &  \hspace{4mm}$3f+1$   & \cmark              & \cmark                   & $f$                                          & $f _{prep}\wedge f _{conf}\wedge f_{exec}$ &  $n$                        &    \hspace{6mm} $0_{exec}$ &   $n$ \\ \bottomrule
    \end{tabular}
    \caption{\textbf{Comparison of fault models in BFT systems. }}
    \label{tab:my-table}
\end{table*}

%% file: sections/03-design.tex
\section{BFT-Centric Partitioning}\label{sec:approach}

In this section, we address the question: 
\textit{How to identify fitting units for different compartments that fail independently}?
We first present several principles that guide decisions about what data and logic should be placed into individual compartments.
We then apply our reasoning to the traditional \ac{PBFT} protocol.

\subsection{Partitioning Principles}
\label{design:principle}

\magic is a customizable approach to designing a multi-enclave BFT architecture. By compartmentalizing safety-sensitive functionalities, we can build a BFT system with improved security and robustness.
The core argument is that this allows us to contain failures within one compartment while the other compartments remain intact even when located on the same replica.
Intel SGX offers this possibility of partitioning a single program into an untrusted environment and multiple secure enclaves and is noticeably characterized by a small TCB compared to other TEEs.
In the following, we propose the principles \circl{$P_1$} - \circl{$P_5$} that can guide decisions on how to separate a BFT protocol into an environment and multiple compartments:


\par
\circl{$P_1$}\textit{\textbf{ Place only safety-critical state and logic into the enclave.}} \magic should maintain safety, even if an attacker is present in the environment of every replica.
Thus, all safety-related logic and state needs to be placed inside an enclave.
If an attacker has compromised the environment, liveness of code running in an enclave cannot be guaranteed.
Therefore, logic that is only relevant for liveness should stay in the untrusted environment.
The separation of liveness results in a reduced TCB and can thus increase the resilience.
Typically, protocol logic must stay in the enclave, while timers, network handling, buffering, static variables, and the output log can remain in the untrusted  environment.

\par
\circl{$P_2$}\textit{\textbf{ Event handlers as decision units.}} 
BFT protocols, as most distributed algorithms are event-driven. They are formulated as a set of event handlers.
An event handler is a function invoked on a timeout or the receipt of a message. Event handlers manipulate shared state and 
may again trigger new messages to be sent.
Event handlers should run until completion in a single compartment. 
Indeed, splitting one event handler into multiple compartments favors complex dependencies that may lead to dependent failures. Also, entering/exiting multiple enclaves during one event handler may be detrimental to performance.
Placing an event handler into a compartment imposes that different enclaves communicate via authenticated messages, as is done between various replicas.

\par
\circl{$P_3$}\textit{\textbf{ Place event handlers that access the same state into one compartment.}}
We restrict the shared state accessed by different compartments to global and static configurations.
To achieve this, event handlers that operate on the same state should be placed in the same compartment. 

We note that a naive application of this principle may easily require all event handlers to be placed in a single enclave.
A more fine-grained design requires carefully separating the data structures used by the algorithm. 
Another possible solution is to duplicate the state and maintain it separately in multiple compartment. However, duplication may lead the system to deviate from the original specification and result in a safety violation (e.g., state divergence). In the case of data structure such as a log of messages, if two compartments rely on each other's messages to trigger an event, one could intentionally omit or withhold messages. 

 \par
 \circl{$P_4$}\textit{\textbf{ Place event handlers including similar logic into one compartment.}}
 Sometimes, event handlers do not share state but execute the same or similar logic. 
 Duplicating such logic in multiple compartments may also duplicate vulnerabilities. 
 We therefore suggest placing these event handlers into one compartment.
\par
\circl{$P_5$}\textit{\textbf{ Place compartment transitions on quorum decisions.}}
When handling individual messages, a faulty sender may easily influence the receiver.
For that reason, BFT protocols typically collect messages from multiple senders and only act upon receiving a super-majority (quorum) of matching messages. These quorum of messages are ideal for compartment transitions since the failure of individual senders 
cannot corrupt the receiving compartment.

\subsection{\magic Principles Applied to \ac{PBFT}}
In the following, we first give some background on the PBFT algorithm and especially the state maintained by the algorithm.
We then discuss how we identify safety critical variables and how we partition event handlers into multiple compartments applying Principles \princ{1} to \princ{5}.

\paragraph{Preliminaries} \ac{PBFT} is regarded as the baseline for almost all published BFT protocols~\cite{ clement2009making,sbft,yin2003separating, kotla2010zyzzyva}. The consensus procedures of PBFT can be described as follows: the agreement is a three-phase protocol with one designated sender process, the \textit{primary} associated with a $view$, decides on the total order of clients' requests in a \prepreparemsg message. All other replicas coordinate through broadcasting \preparemsg messages to validate the primary proposal. Finally, all replicas agree on the total order of requests after receiving a quorum of
\commitmsg messages. These steps are repeated while increasing sequence numbers to order additional requests.
Each replica maintains a copy of an application, e.g., a blockchain.
PBFT uses the ordered requests as input in the execution stage, executes clients' operations, and sends replies. In the case of a blockchain, this execution may entail the creation of a new block.
\ac{PBFT} guarantees two properties: Safety and liveness. Safety implies clients receive correct (linearizable) replies. Liveness ensures that they eventually receive replies, given that the network does not delay messages indefinitely. 
Replicas use timers to detect a faulty primary and trigger the view-change sub-protocol.
Upon agreement of $2f+1$ replicas, the system moves to a new view with a new, dedicated primary.
PBFT discards requests already executed using the checkpointing sub-protocol to prevent the log of messages from growing indefinitely. Periodically, replicas obtain proof that their state is correct by collecting a \textit{certificate} of $2f+1$ \checkpointmsg messages with the same digest and sequence number. In doing so, they can safely remove old entries from the log.

\paragraph{PBFT State}
According to Principle \princ{3}, we need to place not only logic but also state into different compartments.
We base our partitioning on the PBFT pseudocode~\cite{castrothesis}, where each replica is modeled as an I/O automaton.
According to the algorithm,
each replica maintains two message logs, $in$ and $out$, a few variables, and an instance of the application state. The log
$in$ contains received messages and some sent messages that are later needed to validate pre-conditions. The
$out$ contains messages that should be sent.
For variables, we ignore values that can be derived from the input log. For example, the low watermark, 
a variable marking for which sequence numbers messages have been garbage collected, can be derived from \checkpointmsg messages in the log $in$. Similarly, we ignore configuration parameters that stay constant throughout execution and assume they can be safely loaded into enclaves. Examples of such parameters are $n$, the total number of replicas, and public and private keys.

Local variables are mostly related to the execution of requests, e.g. sequence numbers for the last request executed from each client.
Most relevant for the agreement procedure is the \texttt{view} variable, a number used to identify the current primary and ignore all messages sent under previous primaries, i.e. in an earlier view.

\paragraph{Splitting safety and liveness}
Following~\circl{$P_1$}, we start by separating the state into safety-critical variables that should go into the trusted context from liveness variables that may remain in the untrusted environment. 
Following the replica automaton model in PBFT, we argue that the core safety-critical variables consist of the application state and variables related to request execution, the view, and the input log.
We consider the input log $in$ as safety-relevant since even an omission from that log may result in faulty behavior. For example, the omission of sent messages may result in re-sent messages with diverging data, also known as $amnesia$~\cite{bano2020twins}. 
Hence, it is easier to protect the integrity of the log $in$ through the enclave memory than to repeatedly verify the authentication of the included messages.
Contrarily, the $out$ log remains untrusted as it is only relevant to liveness. The same applies to timers and network connections.
Message authentication (i.e., signatures) happens in the enclave before messages are added to the $out$ log. Besides connection handling, we also place the batching of requests into the untrusted environment.

\begin{figure*}[h]
    \input{fig/pbft-workflow.tex}

\caption{PBFT protocol phases and separation of event handlers into compartments. Event handlers are numbered for reference in the text.}
\label{figure:pbftcomp}
\end{figure*}
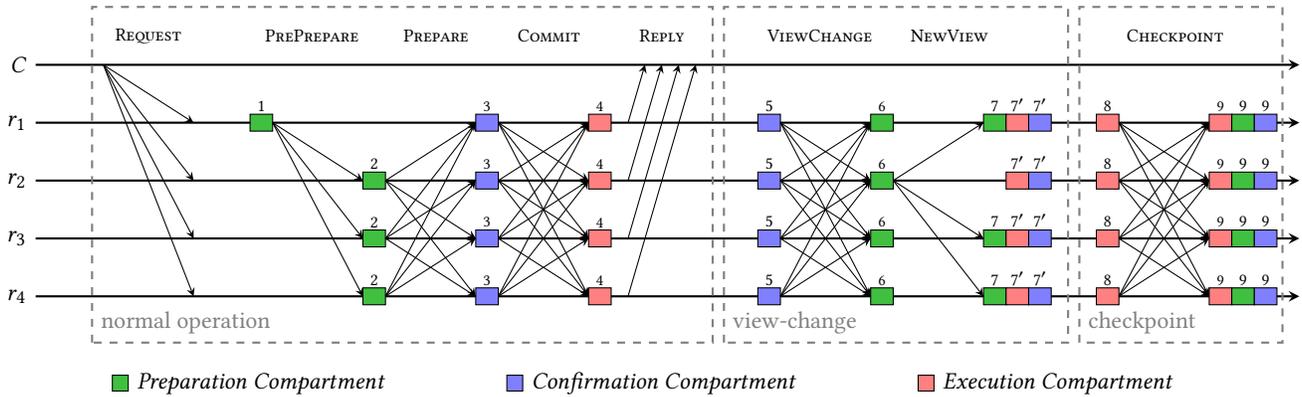

\mypar{Compartments} 
Following the principles \circl{$P_1$} to  \circl{$P_5$}, we partition the \ac{PBFT} protocol into three compartments: 
\begin{itemize}
\item\textit{\preparationC Compartment:} Receives client requests and initializes its order distribution
\item \textit{\confirmationC Compartment}: Confirms that a request was prepared by a
quorum
\item \textit{\executionC Compartment}:  Collects a quorum of confirmations
, executes authenticated requests and sends back the replies to the clients. As this enclave holds the application state, it is also responsible for generating checkpoints. 
\end{itemize}

\noindent
Figure~\ref{figure:pbftcomp} shows the normal operation, view change, and checkpointing subprotocols of \ac{PBFT}. The responsibilities of different compartments are highlighted in different colors.



\paragraph{Separating event handlers}
Following Principle \princ{2}, we separate the event handlers from the PBFT algorithm into three compartments, the \preparationC, \confirmationC, and \executionC Compartment.
Figure~\ref{figure:pbftcomp} shows the different event handlers in the PBFT algorithm. 
Event handler (1) is triggered when receiving a batch of requests from the environment and starts the normal operation.
Event handlers (5) and (8) are triggered through timeouts and start the view-change or checkpointing procedure, respectively. The remaining event handlers are executed every time a respective message is received.

Following principle \circl{$P_3$}, event handlers accessing the same state should be placed in the same compartment. 
However, a naive application of \circl{$P_3$} would result in a single compartment. Indeed, all the identified message handlers need access to the input log $in$ to receive a message; each handler accesses different messages from $in$.
We eliminate this dependency by separating this log into multiple logs containing one message type.

For event handlers (1)-(4) involved in \textbf{normal operation}, we apply Principle \princ{5} and split them according to quorum decisions. We note that no quorum decision happens between event handlers (1) and (2). 
Indeed, the \preparemsg and \prepare{} messages are used together to form a quorum decision in message handler (3).
Thus, if both the enclave sending a \prepreparemsg, and $f$ enclaves of the compartment sending \preparemsg messages are faulty, safety may be violated. Placing these into one compartment makes this dependency clear. 

One problem with this splitting is that \prepreparemsg messages are accessed both by message handlers (2) and (3).
Colocating (2) and (3) would violate \princ{5}.
We therefore duplicate \prepreparemsg{}s in the input log of the \preparationC and \confirmationC Compartment, to avoid shared state (\princ{3}).
We also duplicate \prepreparemsg{}s in the input of the \executionC Compartment.  Thus, the requests are forwarded to this compartment, even if \commitmsg messages only
contain a hash of that request.
 
For the \textbf{checkpointing} subprotocol, we note that a checkpoint message includes a snapshot of the application state. As the application state is already needed to execute client requests in the event handler (4), we place the sending of checkpoint messages (8) with (4) in the \executionC compartment, following \princ{3}.

Finally, the message handler of \checkpointmsg messages~(9) accesses all input logs and deletes old messages.
We therefore opt to duplicate this handler in different compartments.
Following \princ{2} each of the duplicates runs independently without dependencies between the different compartments.
While this forgoes \princ{4} and clearly creates additional performance overhead, we argue that this is acceptable since \checkpointmsg{}s are only performed periodically and lie outside the critical path.

Regarding the \textbf{view change}, we first note that a \viewchangemsg{} includes received \preparemsg and \prepreparemsg messages.
Following \princ{3} the sending of \viewchangemsg~(5) is co-located with the sending of \commitmsg messages~(3) in the confirmation enclave. 
For the handling of \newviewmsg messages, we note that these messages serve two purposes.
First, the \newviewmsg contains a checkpoint, applied similar to the checkpointing subprotocol.
We extract this functionality as event handler~(7') and duplicate it over all compartments, similar to the handling of checkpoints.

Additional to the checkpoints, a \newviewmsg message contains \prepreparemsg{}s, which results in the sending of \preparemsg messages.
Following Principle~\princ{4}, we place the sending~(6) and receiving~(7) of the \newviewmsg in the \preparationC compartment, together with the sending and receiving of \prepreparemsg~(1,2).

One remaining issue is the \texttt{view} variable, which all event handlers use to avoid processing and sending messages belonging to an outdated view.
To avoid merging all event handlers into a single compartment, we instead replicate the view variable.
As in PBFT, the view variable is updated when sending a \viewchangemsg message~(5), or when receiving a valid \newviewmsg message~(7, 7').
Thus, together with the checkpoints in a \newviewmsg message, we update the view in all compartments.


\section{\magic}
In this section, we answer the question of how to maintain correctness when splitting an algorithm.
We first give a detailed description of our split variant of PBFT. 
We then argue why it maintains correctness and report on our effort to verify this correctness formally.

\paragraph{\magic Workflow}We now show a request execution workflow with further details and explain how our design prevents enclaves from tricking each other and ensures correctness and safety. 

\par
1) \textit{Client requests}:  Clients are identified with a pair of keys a $pk_i$, and $sk_i$.
Additionally, each enclave has an individual key pair.
We assume public keys are known to all participants.
At the start of the service, the client first attests to the execution and preparation enclave verifying their genuineness and SGX support. When the attestation is successful, the client provides the execution enclave with a session key $s_{enc}$ 
to encrypt requests and preserve their confidentiality from the untrusted environment and the rest of the enclaves. The encrypted requests are then signed for authentication.
When clients submit corrupted operations, the \executionC Compartment will detect this and execute a no-op instead.

\par
2) \textit{Ordering protocol:} The \preparationC enclave on the primary replica authenticates the request, assigns it a sequence number and then stores it in its input log $in_{prep}$.
After checking the correctness of the request, the preparation enclave creates a \prepreparemsg message, signs it using its private key, and then pushes it into the output log. 
\preparationC enclaves on the backups receive the \prepreparemsg message, verify its correctness, and create and sign a \preparemsg messages. 
\prepreparemsg{}s are also forwarded to the \confirmationC Compartment. 

The \confirmationC Compartment waits for a \preparemsg certificate containing one \prepreparemsg and $2f$ matching \preparemsg messages from different enclaves of type \preparationC Compartment. Then each \confirmationC enclave creates
and signs a \commitmsg. Finally, the \executionC Compartment waits for a quorum of \commitmsg{}s.
Since \commitmsg may only include a hash of the request, the \executionC also receives \prepreparemsg{}s containing the full request.
It then executes requests and sends replies to the clients. 

Following \circl{$P_5$}, we note that in the above, an enclave does not react to a single message from a different compartment, but only reacts to a certificate of $2f+1$ messages.
This ensures that failures in individual enclaves cannot affect other compartments and protects the quorum certificate from any unanticipated changes.



3) \textit{Garbage collection:}  Each compartment keeps a separate private log and executes checkpoints. The \checkpointmsg message originates from the \executionC Compartment, which holds the application state. 
Each compartment erases old message upon receiving $2f+1$ \checkpointmsg{}s. 
Compartments keep the \checkpointmsg{}s and discard messages for sequence numbers before the checkpoint, even if they are received later.

4) \textit{View-change:} 
%
%
As in PBFT, our replicas set a timer on receiving a request from a client. If this timer expires before the request is executed, replicas suspect the primary to be faulty. These timers are managed by the environment but upon suspicion, the \confirmationC Compartment sends the \viewchangemsg message.
This message contains a certificate of $2f+1$ \checkpointmsg{}s and all \preparemsg certificates from $in_{conf}$.
Upon sending this \viewchangemsg, a \confirmationC enclave increases its view. Thus, it will no longer process \preparemsg{}s or send \commit{}s in the old view.
The \preparationC Compartment receives the \viewchangemsg messages, validates them, and sends a \newviewmsg. 

The \newviewmsg in PBFT has three important functions. First, it updates the view and primary on all replicas, second, it distributes a checkpoint, and third, it allows the new primary to resend \prepreparemsg{}s for requests that have been assigned a sequence numbers in the previous view, but are not included in the checkpoint yet.
The new primary needs to create \prepreparemsg messages for the \newviewmsg based on the \preparemsg certificates included in \viewchangemsg{}s. This logic is complex and it is repeated when validating the \newviewmsg in the \preparationC Compartment.
The \confirmationC and \executionC Compartments also receive the \newviewmsg but do not validate the \prepreparemsg{}s included.
They only validate and apply the checkpoint and update their view number, if they did not yet do so.


\paragraph{Correctness}
We first argue informally why our splitting of PBFT is correct. We then report on our formal verification.
We need to argue that our split protocol maintains both liveness and safety if, on $2f+1$ replicas, both the environment and all enclaves are correct. 
If the environment of a replica is correct, \prepreparemsg, \checkpointmsg, and \newviewmsg{}s are forwarded to all compartments at the same time. A replica in \magic thus behaves like a replica in PBFT. Safety and liveness follow from the respective properties in PBFT. 
There is one corner case, namely that a \newviewmsg that contains false \prepreparemsg{}s but is otherwise correct will 
be accepted by the \confirmationC and \executionC Compartment, but not by the \preparationC Compartment.
In this case, the replica may send \commitmsg{}s in the new view but will not send \preparemsg{}s.
We note that a prepare certificate containing $2f$ \preparemsg and one \prepreparemsg message is still needed to send a \commit.
Thus, safety is guaranteed. 

Furthermore, we have to argue that our compartmentalized version of PBFT does maintain safety as long as $2f+1$ enclaves from each compartment-type
are correct. 
In this case, messages may arrive selectively or reordered at the different compartments.
Regarding checkpoints, we note that receiving \checkpointmsg messages in a different order gives the same result.
A replica only handles messages in a fixed window of sequence numbers above the last checkpoint. 
Thus, the replica may not respond to messages in higher sequence numbers after omitting a checkpoint.
This does not endanger safety.

In case the \preparationC Compartment at the primary is faulty, a replica may receive two different \prepreparemsg{}s and 
forward them once to the \preparationC and once to the \confirmationC enclave. However, if $2f+1$ \preparationC enclaves are correct, no two replicas can receive different \preparemsg certificates. Therefore, safety still holds.

Finally, given a faulty environment, the \preparationC enclave may process a \prepreparemsg and send a \preparemsg{} after the \confirmationC enclave has sent a \viewchangemsg.
If that additional \preparemsg is used in a \preparemsg certificate triggering a \commitmsg on some replica, then this replica will include the certificate in its \viewchangemsg. Otherwise, it remains without effect.
BFT protocols are known for their complexity, and some previous protocols are known to contain faults~\cite{abraham2017revisiting,christianPBFTread}. We formally verify that \magic does maintain safety under the complex conditions stated above.
To do that, we proved safety using the Ivy verification tool~\cite{ivy}. 
Ivy is a tool that can verify safety proofs for parametrized models. This gives better confidence in correctness than finite model checkers like TLA+ that only verify the model for fixed parameters and suffer from state explosion.
Since no synchronization between different enclaves on one replica can be ensured in case of a faulty environment, we modeled the different enclaves as individual nodes in our proof.
For deriving the proof, we adjusted an existing proof of PBFT~\cite{ivypbft}. This derivation was surprisingly straightforward and only required minor adjustments to the proof. This gives us additional confidence that our model is correct.\footnote{Formal models are available online: \url{https://github.com/leandernikolaus/splitbft-proofs}.}

\subsection*{Discussions and Extensions}

\textit{Further Compartmentalization.} PBFT is not the only fault-tolerant protocol that can be compartmentalized. The principles and ideas we show apply to other BFT protocols, including streamlined variants such as the recent Hotstuff protocol~\cite{yin2019hotstuff}. 
Besides, the application included in our execution enclave may significantly increase the TCB for certain use cases.
In such cases, further compartmentalization may be applied by 
(i) separating the application in its own enclave and (ii) a more fine-grained partitioning of the application, e.g., applying sharding techniques~\cite{didona2019size}.

\textit{Enclave recovery.} When enclaves are identified to either be corrupted by a memory leak or fail by crashing, we consider rebooting the single affected component. 
Enclaves that possibly store data persistently, e.g., the execution or application enclave, can use the SGX sealing technique for uniquely encrypting the enclave secrets and securely recovering them when rebooting. 
An important issue is \textit{rollback and forking attacks}. In this case, the malicious server can return a correctly checked outdated state or create multiple instances of an enclave.
To detect these attacks, previous works use trusted time or monotonic counter and can be integrated into our design as a defensive technique~\cite{brandenburger2017rollback,bailleu2019speicher}. 

\textit{Denial-of-service and performance attacks.} Note that an enclave is subject to sudden crashes triggered due to a compromised environment. In addition, an exploited enclave could remain unresponsive to messages or delay executing an operation that breaks the protocol liveness or slows down the execution. 
Clients can also harm the performance by consistently sending corrupted operations that will be ordered but ignored during execution. This is no different from a client sending unnecessary but correctly formed requests and can be addressed by rate limiting. Alternatively, the execution enclave can intercept client requests and apply an access control check preventing the ordering of corrupted requests.

\textit{Contribution Above PBFT.}
We show how PBFT can be separated into three independent compartments while maintaining overall correctness. This separation increases resilience,
and especially ensures safety despite failures on all machines and in a fraction of the enclaves.
We also prevent cases where enclaves may consciously break safety, e.g., by duplicating the log entries when necessary. 
The separation is also relevant for performance. Since components are independent, they can be executed in parallel without the need for synchronization.

%% file: fig/pbft-workflow.tex
    \begin{tikzpicture}[>=stealth,x=1cm,y=1.1cm,yscale=0.7,xscale=0.75]
        \stdset{msg label offset=0.5, msg type font=\footnotesize, exec box label font=\scriptsize}
        \initstd
        \process{/srv4}{$r_4$}
        \process{/srv3}{$r_3$}
        \process{/srv2}{$r_2$}
        \process{/srv1}{$r_1$}
        \process{/clt}{$C$}
                
        \advancetime
        \mcast[exec box color=green!50!gray]{/clt}{{/srv1, /srv2, /srv3, /srv4}}{Request}{false}
        \execbox{/srv1}{/srv1}{$1$}{1}
        \advancetime
        \mcast[msg label x offset=-0.1]{/srv1}{{/srv2, /srv3, /srv4}}{PrePrepare}{$2$}
        
        \mtom[exec box color=blue!50!white,msg label x offset=0.1]{{/srv2, /srv3, /srv4}}{{/srv1, /srv2, /srv3, /srv4}}{Prepare}{3}
        \alltoall[exec box color=red!50!white]{{/srv1, /srv2, /srv3, /srv4}}{Commit}{4}
        
        
        \msgx{/srv1}{/clt}{}{0.5}
        \msgx{/srv2}{/clt}{}{0.5}
        \msgx{/srv3}{/clt}{}{0.5}
        \msgx{/srv4}{/clt}{}{0.5}
        \msgtxt{Reply}

        \advancetime[2]
        
        \stdset{exec box color=blue!50!white}
        \execbox{/srv1}{/srv1}{$5$}{1}
        \execbox{/srv2}{/srv2}{$5$}{1}
        \execbox{/srv3}{/srv3}{$5$}{1}
        \execbox{/srv4}{/srv4}{$5$}{1}
        \advancetime
        \alltoall[exec box color=green!50!gray,msg label x offset=-0.1]{{/srv1, /srv2, /srv3, /srv4}}{ViewChange}{6}
        \mcast[exec box color=green!50!gray,msg label x offset=0.2]{/srv2}{{/srv1, /srv3, /srv4}}{NewView}{$7$}
        \stdset{exec box color=red!50!white}
        \execbox{/srv1}{/srv1}{$7'$}{0.4}
        \execbox{/srv2}{/srv2}{$7'$}{0.4}
        \execbox{/srv3}{/srv3}{$7'$}{0.4}
        \execbox{/srv4}{/srv4}{$7'$}{0.4}
        \stdset{exec box color=blue!50!white}
        \execbox{/srv1}{/srv1}{$7'$}{0.8}
        \execbox{/srv2}{/srv2}{$7'$}{0.8}
        \execbox{/srv3}{/srv3}{$7'$}{0.8}
        \execbox{/srv4}{/srv4}{$7'$}{0.8}

        \advancetime

        \stdset{exec box color=red!50!white}
        \execbox{/srv1}{/srv1}{$8$}{1}
        \execbox{/srv2}{/srv2}{$8$}{1}
        \execbox{/srv3}{/srv3}{$8$}{1}
        \execbox{/srv4}{/srv4}{$8$}{1}
        \advancetime

        \alltoall{{/srv1, /srv2, /srv3, /srv4}}{Checkpoint}{9}
        \stdset{exec box color=green!50!gray}
        \execbox{/srv1}{/srv1}{$9$}{0.4}
        \execbox{/srv2}{/srv2}{$9$}{0.4}
        \execbox{/srv3}{/srv3}{$9$}{0.4}
        \execbox{/srv4}{/srv4}{$9$}{0.4}
        \stdset{exec box color=blue!50!white}
        \execbox{/srv1}{/srv1}{$9$}{0.8}
        \execbox{/srv2}{/srv2}{$9$}{0.8}
        \execbox{/srv3}{/srv3}{$9$}{0.8}
        \execbox{/srv4}{/srv4}{$9$}{0.8}

        \advancetime
        \drawtimelines

        \draw[thick, gray, dashed] (1,0.2) node[above right]{normal operation} rectangle (12,6);
        \draw[thick, gray, dashed] (12.2,0.2) node[above right]{view-change} rectangle (18.3,6);
        \draw[thick, gray, dashed] (18.5,0.2) node[above right]{checkpoint} rectangle (22.1,6);

        \node (P) at (1.5,-0.5) [draw,minimum height=0.2,minimum width=0.3, fill=green!50!gray]{};
        \node (Pl) at (P.east)[right]{\textit{Preparation Compartment}};
        \node (C) at ([xshift=2cm]Pl.east) [right,draw,minimum height=0.2,minimum width=0.3, fill=blue!50!white]{};
        \node (Cl) at (C.east)[right]{\textit{Confirmation Compartment}};
        \node (E) at ([xshift=2cm]Cl.east) [right,draw,minimum height=0.2,minimum width=0.3, fill=red!50!white]{};
        \node (El) at (E.east)[right]{\textit{Execution Compartment}};
    \end{tikzpicture}

%% file: sections/04-implemtation.tex
\section{Implementation}\label{sec:implementation}

We implement \magic on top of Themis a Rust-based implementation of PBFT~\cite{rusch2019themis} ( using the \lib{nightly-2021-02-17}).
\par \textit{Trusted enclaves.} 
Enclaves are responsible for preserving the integrity of the pre-selected variables or functions and the confidentiality of requests and replies.
Our initial implementation uses Intel SGX due to its availability, the minimal TCB, and support for multiple TEEs.
Intel SGX implements enclaves that use an isolated reserved part of the DRAM called \ac{EPC},  encrypted and authenticated by the CPU. Two Intel SGX SDKs are available implemented in Rust and C/C++.  
Our implementation uses the rust SDK \lib{Teaclave} v1.1.3~\cite{teaclave}. 
To avoid synchronization primitives inside enclaves, we only allow a single thread to execute in each enclave.

The SDK defines calls between the enclave and the outside world that are part of the application logic. An \textsf{ecall} into the enclave or an \textsf{ocall} into the untrusted side. An essential practice when enclavising an application is minimizing the performance overhead that comes from these transitions ($\approx$ 8,640 cycles~\cite{weisse2017regaining}).

\par
To digitally sign enclave messages, we use the  implementation of 256-bit ED25519 from the \lib{ring} library v0.16.20~\cite{ring}. For authenticating client requests and replies, we use the \lib{HMAC-SHA2} function.

\par \textit{Untrusted broker.} The untrusted host environment holds a shim layer or a broker where enclaves register. The broker is responsible for handling I/O for the enclaves. For sending messages or requiring an I/O operation, enclave handlers request it from the broker by posting ocalls into its queue. Likewise, the broker intercepts incoming messages and sends them to the corresponding enclave using ecalls. The calls into the trusted environment are multi-threaded, i.e., each enclave is associated with a thread that triggers ecalls. The broker expects the data that it needs to send over the network serialized using \lib{serde}~\footnote{\url{https://serde.rs/}}. This layer can be compromised, causing liveness issues or denial of service by not handling events, dropping messages, or communicating the wrong timer to the concerned enclaves. However, confidentiality and integrity are not affected as attempts to tamper with the data are detected.


\begin{table}[]

    \begin{tabular}{@{}lp{1.2cm}p{1cm}p{1cm}|p{1.5cm}@{}}
    \toprule
                       & \textbf{Shared types}  &\textbf{Logic} & \textbf{Total LOC} & \textbf{Binary Size (MB)} \\ \midrule
    Preparation Enc.   & 2430  & 487 & 2917 & 1.1               \\
    Confirmation Enc.  & 2430  & 458 & 2888 & 1.1               \\
    Execution Enc.     & 2430& 579 & 3009 & 1.2                \\
    Untrusted Env.     &  &  & 12565  & ---                 \\
    Trusted Counter    &  &  & 439   & 0.524                \\ \bottomrule
    \end{tabular}
    \caption{\textbf{TCB sizes for all enclaves}}
\label{table2}
\end{table}

\mypar{Analysis}
As there is a correlation between the amount of code and the likelihood of vulnerabilities or defects, we analyze our software in terms of lines of code (using the \lib{tokei} utility) and enclave size.
We show the code that executes within each enclave and the untrusted infrastructure. Table \ref{table2} shows the resulting LOC.
The table shows separate line numbers for the type definitions and data structures used in all enclaves and the logic unique to the enclave.
While the Teaclave SGX SDK, \lib{serde}, and ring dependencies are included in the TCB, they are not included in the numbers presented. The untrusted line consists of the broker layer, the communication handling between replicas, and networking (i.e., sockets).
For comparison, we also report the LOC of a Rust implementation of the trusted counter, as used in hybrid systems.
The execution enclave code mostly depends on the application and how the developer decides to engineer it. In our case, the LOC of the execution enclave includes the key-value store. 
While these numbers do not reveal the complexity of the implementation, it gives us an impression of the TCB. Indeed, we show that individual enclaves are significantly smaller than a complete application.
An attacker who targets a non-split application has a larger attacker surface to explore than a split BFT application.

%% file: sections/05-evaluation.tex
\section{Evaluation}\label{sec:evaluation}


We evaluated \magic to get an answer to the following questions:
\begin{itemize}
\item What is the overhead of \magic compared to a common PBFT implementation? 
\item How does \magic perform with realistic applications?
\item What is the overhead of the SGX enclaves?
\end{itemize}

\mypar{Experimental Setup}  We deploy \magic on a cluster
of four SGX-enabled Azure VMs Standard DC4s\_v2 (4 vcpus, 16 GiB memory), each equipped with an Intel Xeon E-2288G
CPU comprising four cores running at 3.7 GHz without HyperThreading connected via 40 Gb switched Ethernet and 16 GB of memory.
 The software
environment of the machines includes a Linux Ubuntu 18.04 with kernel 5.4.0-1074-azure and the Intel SGX SDK in version 2.16. 
A VM machine Standard\_D8s\_v3 (8 vcpus) is dedicated to running the client implementation that
generates the workload. The VMs are all located in the same region \textit{West Europe}. 




\begin{figure*}[bt]
\begin{subfigure}

\hspace{-1.74cm}
\subfigure{\includegraphics[width=8.09cm]{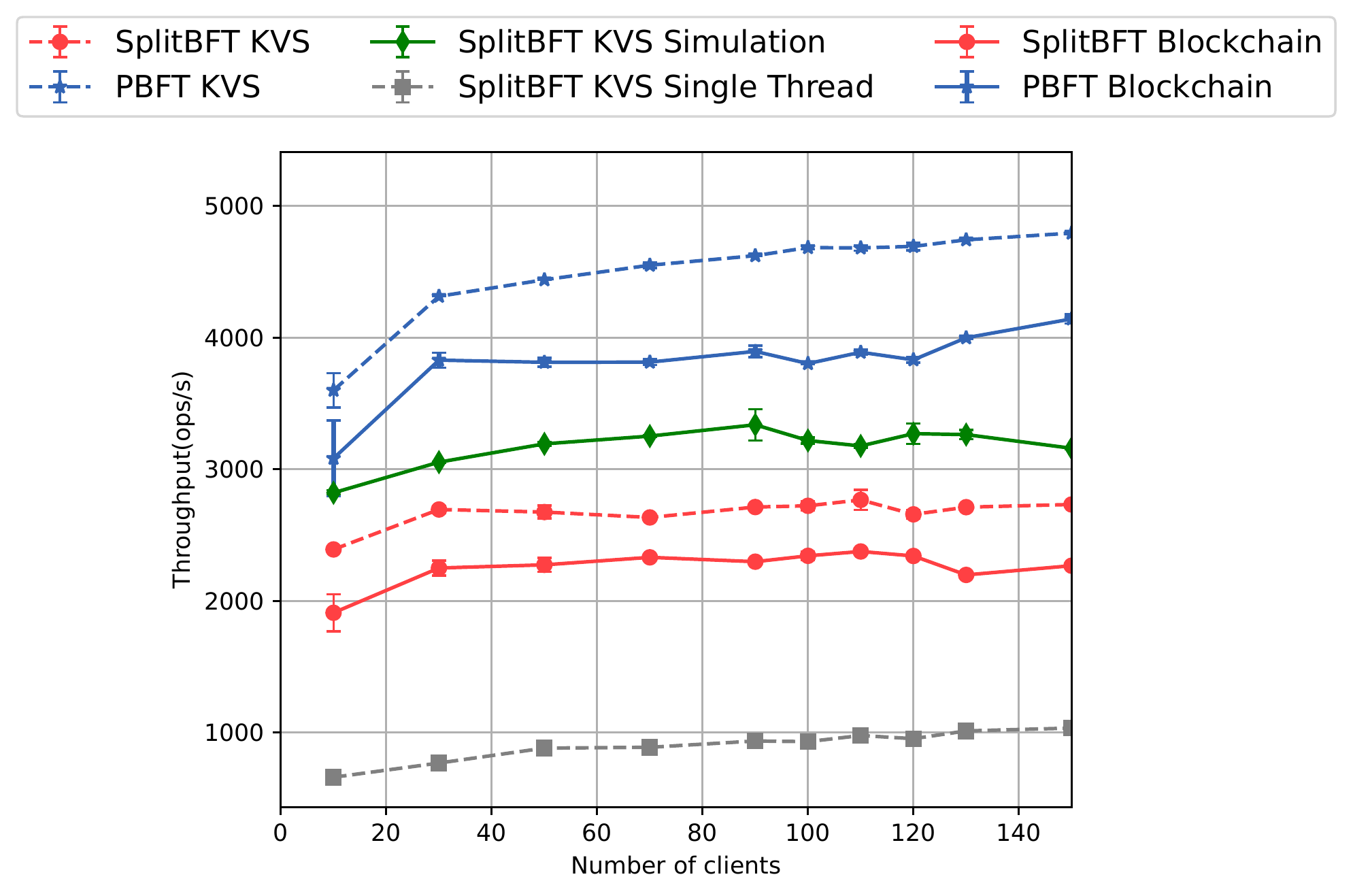}} 
\hspace{-0.45cm}
\subfigure{\includegraphics[width=5.47cm]{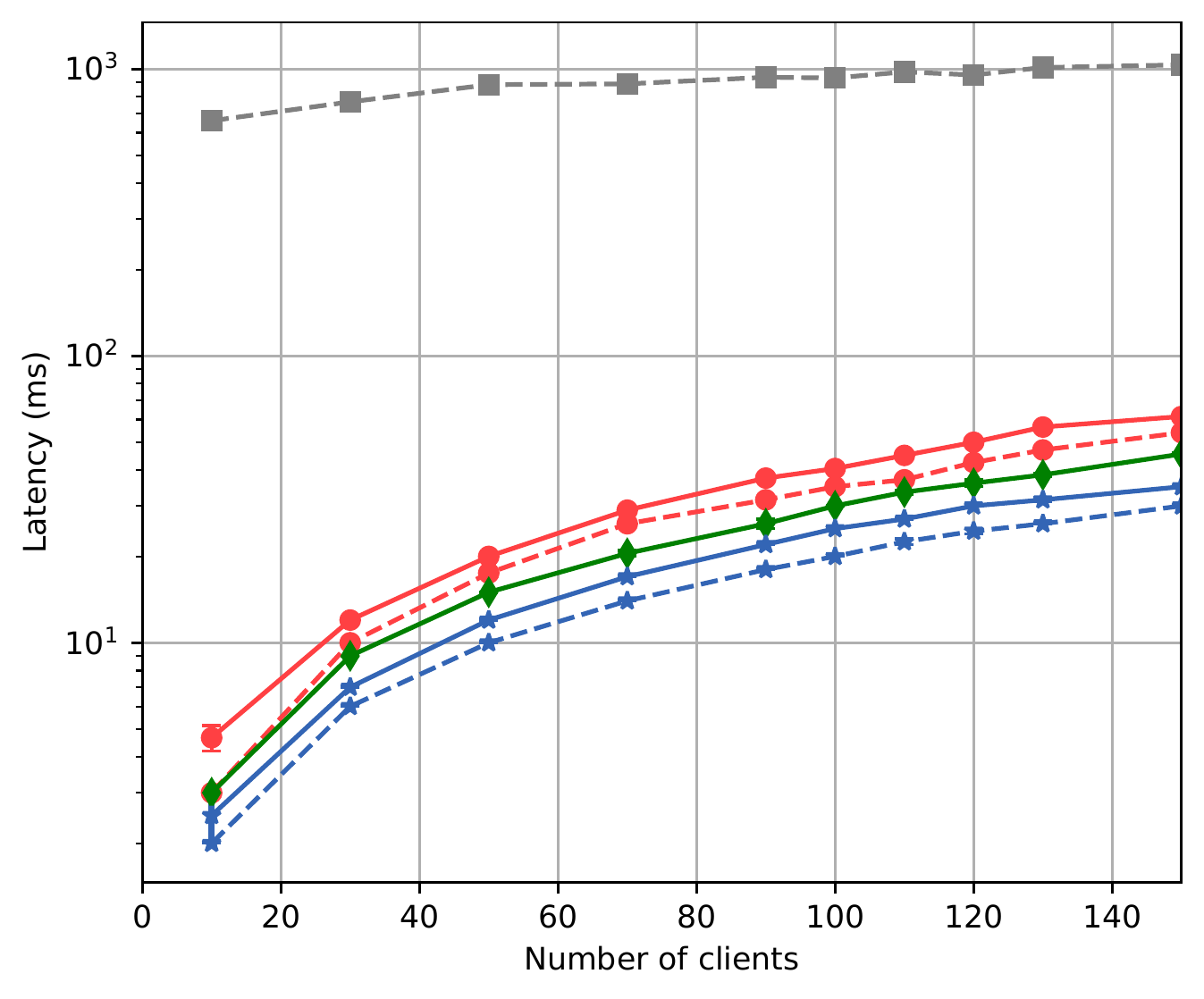}} 
\hspace{-1.1cm}
 \captionsetup[table]{font=small,skip=0pt}
\vspace{-0.5cm}
\captionsetup{labelformat=empty,labelsep=none}
\renewcommand{\thefigure}{(a)}
\caption{\normalfont{(a) Not Batched}}
\label{fig:notbatchedrps}
\end{subfigure}

\begin{subfigure}

\hspace{0.1cm}
\subfigure{\includegraphics[width=5.8cm]{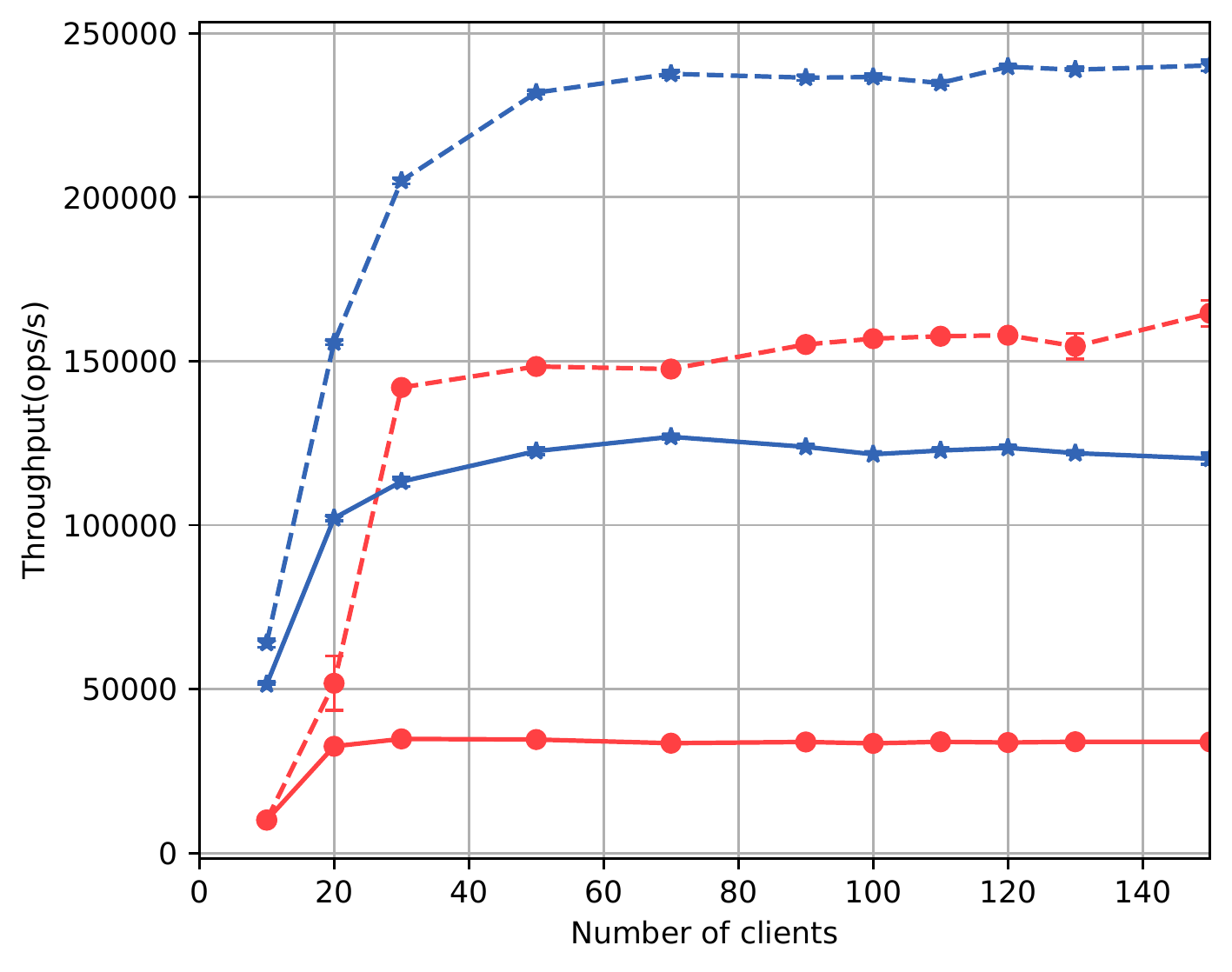}} 
\hspace{1.1cm}
\subfigure{\includegraphics[width=5.5cm]{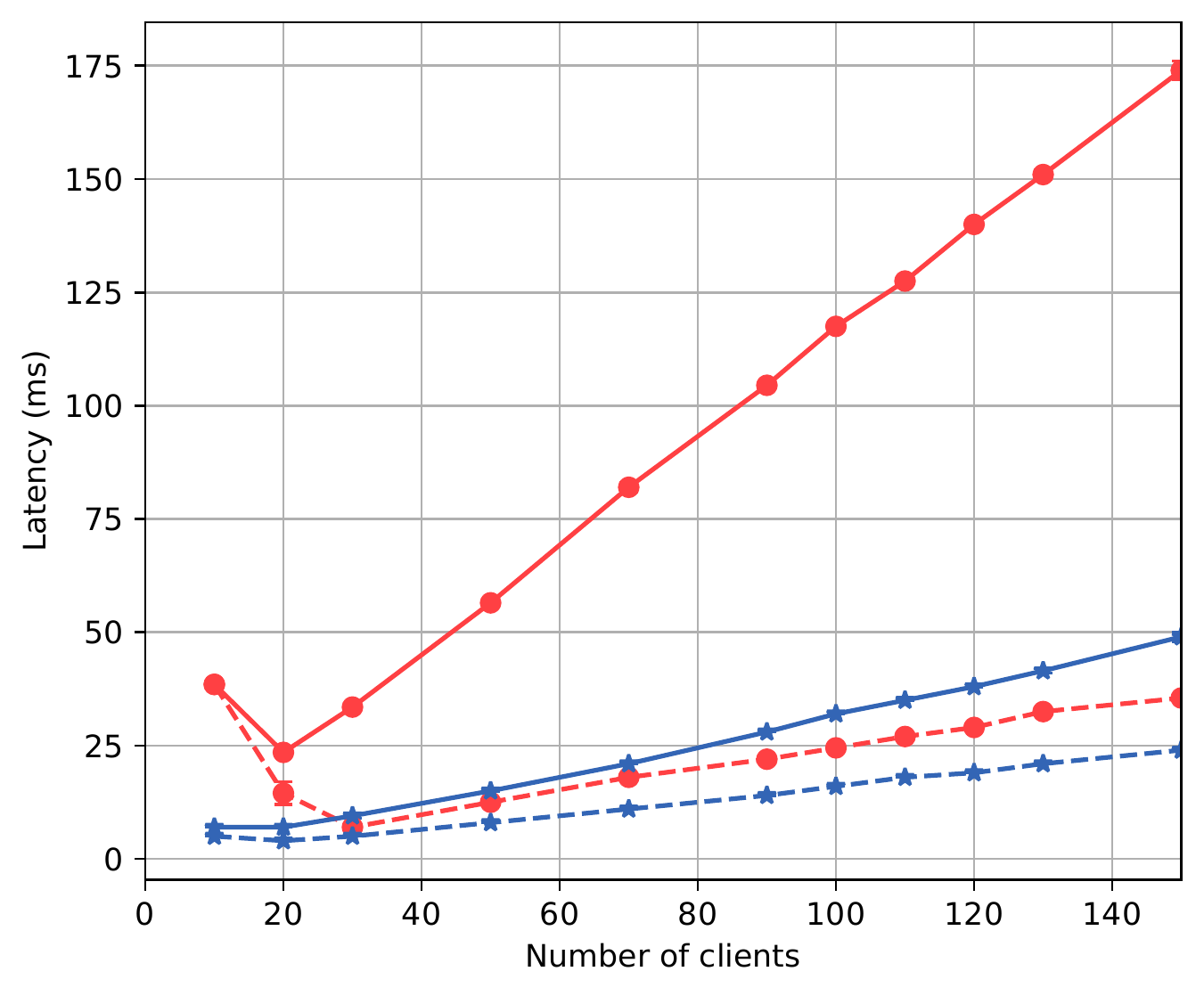}} 
\captionsetup[table]{font=small,skip=0pt}
\vspace{-0.5cm}
\captionsetup{labelformat=empty,labelsep=none}
\renewcommand{\thefigure}{(b)}
\addtocounter{figure}{-1}
\caption{\normalfont{(b) Batched}}
\label{plot:batchedrps}
\end{subfigure}
\addtocounter{figure}{-1}
\caption{Throughput~(ops/s) and latency~(ms) for \magic and PBFT without batching and with 200 batches using two applications: a blockchain and a key-value store~(KVS). }
\label{fig:rpslatency}
\end{figure*}

\begin{figure}
\centering
\includegraphics[width=0.43\textwidth]{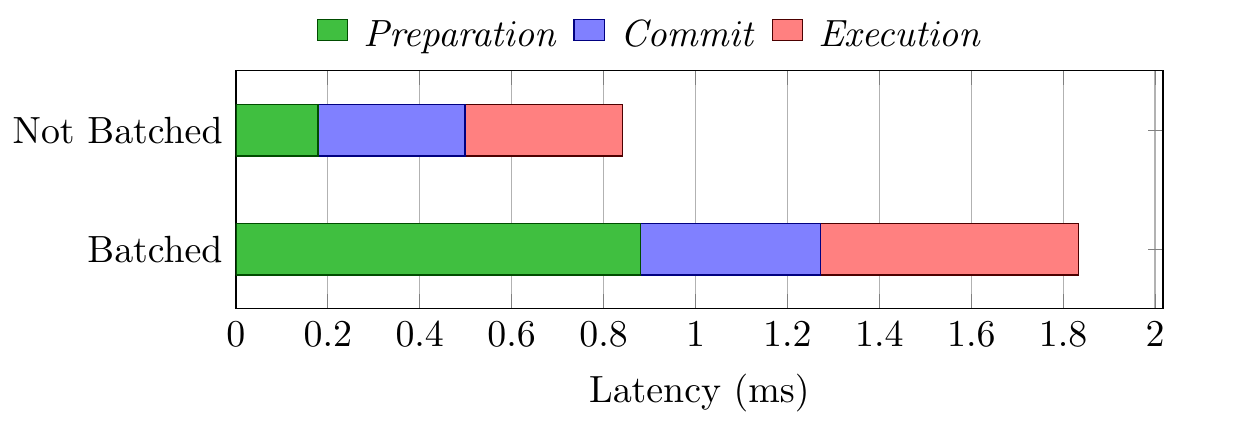}

\caption{Average latency for ecalls done during the processing of one request/batch in different compartments for 40 clients. Measurements are taken on the leader using the KVS application.}
\label{figure:ecall}
\end{figure}




\mypar{Configuration and baseline} We chose PBFT as a general baseline to evaluate the performance of \magic with batching and without batching using a payload of 10 bytes and reply size of 10 bytes. 
\magic uses a dedicated thread for each enclave, which performs ecalls, and an additional thread running the event loop.
We configure PBFT to use a pool of 4 worker threads using the work stealing thread pool from the \textsc{tokio} library. Instead, \magic uses regular OS threads that are more suited for the enclaves development.

As SplitBFT, our PBFT implementation uses HMACs to authenticate client requests and responses but signatures for messages between replicas.
In our PBFT implementation, networking and message authentication are parallelized, but the core protocol is not.
We target two custom applications as use-cases: a key-value store and a blockchain. Clients constantly issue synchronous requests in all our measurements and measure the time it takes to collect the replies. We report the latency and throughput based on these measurements and save the average of five runs.
The blockchain application creates blocks of five messages in the execution enclave and writes them using an \lib{ocall} into the untrusted memory to be stored and encrypted persistently. For that, we use the\lib{ sgx\_tprotected\_fs} crate that allows secure I/O operations. Our throughput and latency measurements evaluate a \textsc{PUT} operation that updates the entries. 
To better understand the overhead introduced by \magic we also measured the latency for different ecalls done on our enclaves.
Our evaluation shows that multithreading significantly reduces the overhead of \magic and that the overhead due to enclave transitions can be amortized over request batches.

\mypar{Throughput and Latency Without Batching} 
Figure~\ref{fig:rpslatency} ~\ref{fig:notbatchedrps} shows the throughput and latency without batching. 
We see that \magic reaches about 43\%-74\% of the throughput of PBFT for the key-value store and 38\%-59\% for the blockchain application. The key-value store application performs up to 33\% better than the blockchain application due to the additional I/O operation and encryption when writing a block persistently.

Compared to the baseline, \magic adds performance overhead due to many reasons, including: (i) enclave transitions, (ii) data copying in and out of enclaves, and (iii) added serialization and de-serialization of protocol messages and client requests.  From the results, we observe that measuring \magic with a single thread performing all ecalls reduces the performance significantly.
To better understand the overhead, we evaluate \magic running enclaves in simulation mode. As the simulation mode omits costly enclave transitions, results suggest that (i) enclave transitions cause 20\% of the overhead. 
To further analyze the overhead, we measure execution times of the ecalls performed in different enclaves during normal request processing. 
Figure~\ref{figure:ecall} shows the average time spent in different enclaves during the processing of one request. 
All ecalls sum up to 841 $\mu$s. Thus, if a single thread is performing all ecalls, a maximum throughput of $\approx$1190 rps could be reached.
Without batching, ecalls to the \executionC compartment have the longest latency, with a total of 343$\mu$s.
In multithreaded \magic, a single thread performs all ecalls to the \executionC compartment.
This thread thus cannot process more than 2900rps.
We see that the throughput in Figure~\ref{fig:notbatchedrps} comes close to these theoretical upper limits.
Thus, the overhead for unbatched \magic is due to ecalls to the \executionC compartment. 
The added overhead in the blockchain application is also located in the \executionC compartment and therefore effects the overall throughput.


\mypar{Throughput and Latency With Batching} 
Figure~\ref{fig:rpslatency}~\ref{plot:batchedrps}  show results for \magic and PBFT, when client requests are processed as batches. This experiment allows each client to have 40 outstanding requests in parallel. In both systems, we create batches on either receiving 200 requests or expiration of a 10ms timeout. The results show that \magic reaches $\approx$ 64\% the throughput of PBFT for the key-value store and 55\% for the blockchain application.  \magic key-value store performs better than the blockchain application with up to a 4.6$\times$ more throughput. Indeed, the execution enclave performs one ocall for each block (5 requests), while in the case of the key-value store, we only perform one ocall on executing each batch. 

From Figure~\ref{figure:ecall}, we see that the ecalls to the \preparationC and \executionC compartment are significantly longer now.
Ecalls to the \confirmationC compartment are similar to the unbatched mode since this compartment only handles a hash of the request batch. 
Here, ecalls to the \preparationC compartment are the longest. 
These ecalls give a theoretical upper limit of $\approx$
227k operation per second.  
The long ecalls compared to the non batched mode are due to the authentication verification of a batch of client requests instead of a single request and the copy in/out of the enclave. 
For the \executionC,
requests need to be un-marshaled and executed. Responses need to be authenticated and copied out of the enclave.
The last is also a good example of how enclaves add additional overhead. 
All responses are collected, marshaled, and passed out of the enclave to avoid multiple enclave transitions (ocalls).
The collection then needs to be split and sent to individual clients.
On the other hand, in PBFT, marshaling and authentication of different responses are performed concurrently by all threads.

%% file: sections/06-relatedwork.tex
\section{Related Work}

\mypar{Using TEEs on BFT protocols and Blockchains}
A number of BFT systems explored the use of TEEs to isolate a small fraction of the system functionality thereby establishing a hybrid fault model, where the protected part is excluded from the Byzantine fault assumption and can only fail by crashing~\cite{behl2017hybrids, minbft, cheapbft,liu2018scalable}. 
In particular, they aim at reducing the degree of replication by achieving non-equivocation. Clement et al.~\cite{clement2012limited} show that non-equivocation is not enough to decrease the number of required replicas unless it is coupled with transferable authentication. 

MinBFT~\cite{minbft}, CheapBFT~\cite{cheapbft} and Hybster~\cite{behl2017hybrids} assume a trusted subsystem where replicas sign messages using a monotonic counter to address equivocation and thus reduce the fault requirement to $2f+1$. Damysus~\cite{decouchant2022damysus} addresses the case of streamlined BFT protocols such as HotStuff, proposing two trusted services that improve the resilience and reduce the communication rounds as in Hybrid protocols. 
These services record additional information relevant to blocks to guarantee nodes cannot lie about the last prepared blocks.
Recent Avocado places a crash tolerant replica into a TEE, to tolerate ensure confidentiality and integrity despite an attacker present on all nodes, but assumes that TEEs remain correct.

Fairy~\cite{stathakopoulou2021adding} leverages TEEs as a layer on top of an ordering service to add fairness when executing client's requests.
Troxy ~\cite{troxy} uses a TEE to intercept client requests and replies as a proxy layer, resulting in removing the client-side library functionality and making the use of BFT transparent while improving performance. 
However, these hybrid protocols as well as the work of Clement et al. assume achieving equivocation entails the TEE only fails by crashing, an assumption that comes with a high confidence in the protected code, especially with increased TCB. 
A recent line of research aims to leverage the security guarantees of TEEs in blockchain and BFT. 
CCF~\cite{russinovich2019ccf} implements a consortium-based blockchain, assumes that TEE can deviate from the utilized protocol. Therefore, the services record enough signed evidence to attribute the TEE. Ekiden~\cite{cheng2019ekiden} uses TEE-backed smart contracts to preserve confidentiality in blockchain applications. 

Another line of Blockchain and cryptocurrency solutions~\cite{sasson2014zerocash, wang2022zero} rely on Zero-knowledge proof as an encryption scheme to protect their users' privacy which is application-specific and requires a large amount of computation power. 

\mypar{Separating Execution Replica}
Previous research focused on the separation of the responsibility of execution replicas~\cite{yin2003separating, cheng2019ekiden}.  Yin et al.~\cite{yin2003separating} give a system that runs the agreement and execution on separate clusters, preventing execution replicas from leaking confidentiality through a privacy firewall. This also brings the benefit of reducing the number of execution replicas. Other works such as Spare~\cite{distler2011spare}, TwinBFT~\cite{dettoni2013byzantine} rely on separation using virtualization, facilitating recovery, or as a trusted separate component. Hyperledger Fabric~\cite{androulaki2018hyperledger} separates the ordering and execution allowing it to occur in a separate processes which gives performance advantages.
\mypar{Alternative fault models} Many proposed alternative faults models to reduce the complexity or require fewer replicas, e.g., XFT~\cite{liu16osdi}, FaB~\cite{martin2006fast}, UpRight~\cite{clement09sosp}.
These works do not target improving the resilience to byzantine faults as in \magic and focus mainly on performance and scalability.
UpRight~\cite{clement09sosp} and Visigoth~\cite{porto15eurosys} enable different thresholds for safety and liveness. 

\mypar{Resilience and Robustness} A common goal in BFT research is to further increase the robustness and the resilience of a protocol, as preserving the initial fault model seems hard to achieve in practical scenarios. 
Flexible BFT~\cite{xiang2021strengthened} aim is to prevent a fraction of faults besides byzantine, assuming an alive-but-corrupt replica, which may deviate from breaking safety however will not try to break liveness. 
BFT2F explores the design space beyond $f$ failures~\cite{li2007beyond}. When no more than $f$ replicas fail, it preserves the same guarantees as \ac{PBFT}. With more than $f$, it prohibits certain kinds of safety violations.
While these works improves the resilience to some degree, some rely on certain timing assumptions. Furthermore, none leverage the security of TEEs to increase the resilience to byzantine faults.

\mypar{Partitioning}
Partitioning applications is not a new concept and has been explored in different scenarios.
Whittaker et al.~\cite{whittaker2020scaling} presents compartmentalized Paxos to improve the scalability and performance. It separates the logic of the application based on the identified bottleneck. To our knowledge, \magic is the first protocol to leverage compartmentalization based on TEEs to increase resilience. 


%% file: sections/07-conclusion.tex
\section{Conclusion}

\magic{} introduces compartmentalization based on \ac{TEE} to Byzantine fault tolerance.
While our approach is neutral to the availability guarantees of BFT, integrity and confidentiality are substantially strengthened.
\magic{} is especially useful in cloud-based deployments where resources and availability are the main concerns of a provider, and the provider can, to some extent, be relieved from concerns regarding the integrity and confidentiality of the hosted code and data. 
This becomes particularly evident in the context of \ac{BaaS}, where the central role of the cloud providers contradicts the decentralization and fault tolerance demands of permission blockchains.
The performed experiments highlight the approach's feasibility but make the additional overhead of switching between different TEE compartments visible. 
While \magic{} exercises the compartmentalization of PBFT using \ac{SGX}, the approach can be transferred to other BFT protocols that feature quorum decisions as part of the agreement and other trusted execution technology that enables fine-grained trusted execution preferable at the process level.

%% file: main.bbl

\begin{thebibliography}{65}


\ifx \showCODEN    \undefined \def \showCODEN     #1{\unskip}     \fi
\ifx \showDOI      \undefined \def \showDOI       #1{#1}\fi
\ifx \showISBNx    \undefined \def \showISBNx     #1{\unskip}     \fi
\ifx \showISBNxiii \undefined \def \showISBNxiii  #1{\unskip}     \fi
\ifx \showISSN     \undefined \def \showISSN      #1{\unskip}     \fi
\ifx \showLCCN     \undefined \def \showLCCN      #1{\unskip}     \fi
\ifx \shownote     \undefined \def \shownote      #1{#1}          \fi
\ifx \showarticletitle \undefined \def \showarticletitle #1{#1}   \fi
\ifx \showURL      \undefined \def \showURL       {\relax}        \fi
\providecommand\bibfield[2]{#2}
\providecommand\bibinfo[2]{#2}
\providecommand\natexlab[1]{#1}
\providecommand\showeprint[2][]{arXiv:#2}

\bibitem[Abraham et~al\mbox{.}(2017)]%
        {abraham2017revisiting}
\bibfield{author}{\bibinfo{person}{Ittai Abraham}, \bibinfo{person}{Guy
  Golan-Gueta}, \bibinfo{person}{Dahlia Malkhi}, \bibinfo{person}{Lorenzo
  Alvisi}, \bibinfo{person}{Ramakrishna Kotla}, {and}
  \bibinfo{person}{Jean-Philippe Martin}.} \bibinfo{year}{2017}\natexlab{}.
\newblock \showarticletitle{Revisiting Fast Practical Byzantine Fault
  Tolerance}.
\newblock  (\bibinfo{year}{2017}).
\newblock
\urldef\tempurl%
\url{https://doi.org/10.48550/arXiv.1712.01367}
\showDOI{\tempurl}
\showeprint{arXiv:hep-ph/9609357}


\bibitem[Amazon(2022)]%
        {amazonblockchain}
\bibfield{author}{\bibinfo{person}{Amazon}.} \bibinfo{year}{2022}\natexlab{}.
\newblock \bibinfo{booktitle}{\emph{Amazon Managed Blockchain}}.
\newblock
\urldef\tempurl%
\url{https://aws.amazon.com/managed-blockchain/}
\showURL{%
Retrieved Mai 9, 2022 from \tempurl}


\bibitem[Androulaki et~al\mbox{.}(2018)]%
        {androulaki2018hyperledger}
\bibfield{author}{\bibinfo{person}{Elli Androulaki}, \bibinfo{person}{Artem
  Barger}, \bibinfo{person}{Vita Bortnikov}, \bibinfo{person}{Christian
  Cachin}, \bibinfo{person}{Konstantinos Christidis}, \bibinfo{person}{Angelo
  De~Caro}, \bibinfo{person}{David Enyeart}, \bibinfo{person}{Christopher
  Ferris}, \bibinfo{person}{Gennady Laventman}, \bibinfo{person}{Yacov
  Manevich}, \bibinfo{person}{Srinivasan Muralidharan}, \bibinfo{person}{Chet
  Murthy}, \bibinfo{person}{Binh Nguyen}, \bibinfo{person}{Manish Sethi},
  \bibinfo{person}{Gari Singh}, \bibinfo{person}{Keith Smith},
  \bibinfo{person}{Alessandro Sorniotti}, \bibinfo{person}{Chrysoula
  Stathakopoulou}, \bibinfo{person}{Marko Vukoli\'{c}},
  \bibinfo{person}{Sharon~Weed Cocco}, {and} \bibinfo{person}{Jason Yellick}.}
  \bibinfo{year}{2018}\natexlab{}.
\newblock \showarticletitle{Hyperledger Fabric: A Distributed Operating System
  for Permissioned Blockchains}. In \bibinfo{booktitle}{\emph{Proceedings of
  the Thirteenth EuroSys Conference}} (Porto, Portugal)
  \emph{(\bibinfo{series}{EuroSys '18})}. \bibinfo{publisher}{ACM},
  \bibinfo{address}{New York, NY, USA}, Article \bibinfo{articleno}{30},
  \bibinfo{numpages}{15}~pages.
\newblock
\showISBNx{9781450355841}
\urldef\tempurl%
\url{https://doi.org/10.1145/3190508.3190538}
\showDOI{\tempurl}


\bibitem[Arnautov et~al\mbox{.}(2016)]%
        {arnautov2016scone}
\bibfield{author}{\bibinfo{person}{Sergei Arnautov}, \bibinfo{person}{Bohdan
  Trach}, \bibinfo{person}{Franz Gregor}, \bibinfo{person}{Thomas Knauth},
  \bibinfo{person}{Andre Martin}, \bibinfo{person}{Christian Priebe},
  \bibinfo{person}{Joshua Lind}, \bibinfo{person}{Divya Muthukumaran},
  \bibinfo{person}{Dan O'Keeffe}, \bibinfo{person}{Mark~L. Stillwell},
  \bibinfo{person}{David Goltzsche}, \bibinfo{person}{David Eyers},
  \bibinfo{person}{R\"{u}diger Kapitza}, \bibinfo{person}{Peter Pietzuch},
  {and} \bibinfo{person}{Christof Fetzer}.} \bibinfo{year}{2016}\natexlab{}.
\newblock \showarticletitle{SCONE: Secure Linux Containers with Intel SGX}. In
  \bibinfo{booktitle}{\emph{Proceedings of the 12th USENIX Conference on
  Operating Systems Design and Implementation}} (Savannah, GA, USA)
  \emph{(\bibinfo{series}{OSDI'16})}. \bibinfo{publisher}{USENIX Association},
  \bibinfo{address}{USA}, \bibinfo{pages}{689--703}.
\newblock
\showISBNx{9781931971331}


\bibitem[Bahsoun et~al\mbox{.}(2015)]%
        {bahsoun2015making}
\bibfield{author}{\bibinfo{person}{Jean-Paul Bahsoun}, \bibinfo{person}{Rachid
  Guerraoui}, {and} \bibinfo{person}{Ali Shoker}.}
  \bibinfo{year}{2015}\natexlab{}.
\newblock \showarticletitle{Making BFT Protocols Really Adaptive}. In
  \bibinfo{booktitle}{\emph{2015 IEEE International Parallel and Distributed
  Processing Symposium}}. \bibinfo{publisher}{IEEE},
  \bibinfo{address}{Hyderabad, India}, \bibinfo{pages}{904--913}.
\newblock
\urldef\tempurl%
\url{https://doi.org/10.1109/IPDPS.2015.21}
\showDOI{\tempurl}


\bibitem[Bailleu et~al\mbox{.}(2019)]%
        {bailleu2019speicher}
\bibfield{author}{\bibinfo{person}{Maurice Bailleu}, \bibinfo{person}{J\"{o}rg
  Thalheim}, \bibinfo{person}{Pramod Bhatotia}, \bibinfo{person}{Christof
  Fetzer}, \bibinfo{person}{Michio Honda}, {and} \bibinfo{person}{Kapil
  Vaswani}.} \bibinfo{year}{2019}\natexlab{}.
\newblock \showarticletitle{SPEICHER: Securing LSM-Based Key-Value Stores Using
  Shielded Execution}. In \bibinfo{booktitle}{\emph{Proceedings of the 2019
  USENIX Conference on Usenix Annual Technical Conference}} (Boston, MA, USA)
  \emph{(\bibinfo{series}{FAST'19})}. \bibinfo{publisher}{USENIX Association},
  \bibinfo{address}{USA}, \bibinfo{pages}{173--190}.
\newblock
\showISBNx{9781931971485}


\bibitem[Bano et~al\mbox{.}(2020)]%
        {bano2020twins}
\bibfield{author}{\bibinfo{person}{Shehar Bano}, \bibinfo{person}{Alberto
  Sonnino}, \bibinfo{person}{Andrey Chursin}, \bibinfo{person}{Dmitri
  Perelman}, {and} \bibinfo{person}{Dahlia Malkhi}.}
  \bibinfo{year}{2020}\natexlab{}.
\newblock \showarticletitle{Twins: White-Glove Approach for BFT Testing}.
\newblock  (\bibinfo{year}{2020}).
\newblock
\urldef\tempurl%
\url{https://doi.org/10.48550/arXiv.2004.10617}
\showDOI{\tempurl}
\showeprint{arXiv:2004.10617}


\bibitem[Baudet et~al\mbox{.}(2019)]%
        {libra}
\bibfield{author}{\bibinfo{person}{Mathieu Baudet}, \bibinfo{person}{Avery
  Ching}, \bibinfo{person}{Andrey Chursin}, \bibinfo{person}{George Danezis},
  \bibinfo{person}{Fran{\c{c}}ois Garillot}, \bibinfo{person}{Zekun Li},
  \bibinfo{person}{Dahlia Malkhi}, \bibinfo{person}{Oded Naor},
  \bibinfo{person}{Dmitri Perelman}, {and} \bibinfo{person}{Alberto Sonnino}.}
  \bibinfo{year}{2019}\natexlab{}.
\newblock \showarticletitle{State machine replication in the libra blockchain}.
\newblock \bibinfo{journal}{\emph{The Libra Assn., Tech. Rep}}
  (\bibinfo{year}{2019}).
\newblock


\bibitem[Behl et~al\mbox{.}(2017)]%
        {behl2017hybrids}
\bibfield{author}{\bibinfo{person}{Johannes Behl}, \bibinfo{person}{Tobias
  Distler}, {and} \bibinfo{person}{R\"{u}diger Kapitza}.}
  \bibinfo{year}{2017}\natexlab{}.
\newblock \showarticletitle{Hybrids on Steroids: SGX-Based High Performance
  BFT}. In \bibinfo{booktitle}{\emph{Proceedings of the Twelfth European
  Conference on Computer Systems}} (Belgrade, Serbia)
  \emph{(\bibinfo{series}{EuroSys '17})}. \bibinfo{publisher}{ACM},
  \bibinfo{address}{New York, NY, USA}, \bibinfo{pages}{222--237}.
\newblock
\showISBNx{9781450349383}
\urldef\tempurl%
\url{https://doi.org/10.1145/3064176.3064213}
\showDOI{\tempurl}


\bibitem[Ben~Sasson et~al\mbox{.}(2014)]%
        {sasson2014zerocash}
\bibfield{author}{\bibinfo{person}{Eli Ben~Sasson}, \bibinfo{person}{Alessandro
  Chiesa}, \bibinfo{person}{Christina Garman}, \bibinfo{person}{Matthew Green},
  \bibinfo{person}{Ian Miers}, \bibinfo{person}{Eran Tromer}, {and}
  \bibinfo{person}{Madars Virza}.} \bibinfo{year}{2014}\natexlab{}.
\newblock \showarticletitle{Zerocash: Decentralized Anonymous Payments from
  Bitcoin}. In \bibinfo{booktitle}{\emph{2014 IEEE Symposium on Security and
  Privacy}}. \bibinfo{publisher}{IEEE}, \bibinfo{address}{San Jose,
  California}, \bibinfo{pages}{459--474}.
\newblock
\urldef\tempurl%
\url{https://doi.org/10.1109/SP.2014.36}
\showDOI{\tempurl}


\bibitem[Berger et~al\mbox{.}(2021)]%
        {christianPBFTread}
\bibfield{author}{\bibinfo{person}{Christian Berger}, \bibinfo{person}{Hans~P.
  Reiser}, {and} \bibinfo{person}{Alysson Bessani}.}
  \bibinfo{year}{2021}\natexlab{}.
\newblock \showarticletitle{Making Reads in BFT State Machine Replication Fast,
  Linearizable, and Live}. In \bibinfo{booktitle}{\emph{2021 40th International
  Symposium on Reliable Distributed Systems (SRDS)}}.
  \bibinfo{publisher}{IEEE}, \bibinfo{address}{Chicago, IL, USA},
  \bibinfo{pages}{1--12}.
\newblock
\urldef\tempurl%
\url{https://doi.org/10.1109/SRDS53918.2021.00010}
\showDOI{\tempurl}


\bibitem[Blanchemain(2018)]%
        {azure}
\bibfield{author}{\bibinfo{person}{Ga\"{e}l Blanchemain}.}
  \bibinfo{year}{2018}\natexlab{}.
\newblock \bibinfo{booktitle}{\emph{Azure BaaS}}.
\newblock
\urldef\tempurl%
\url{https://docs.nethereum.com/en/latest/azure/set-up-blockchain-on-azure/}
\showURL{%
Retrieved July 7, 2021 from \tempurl}


\bibitem[Brandenburger et~al\mbox{.}(2017)]%
        {brandenburger2017rollback}
\bibfield{author}{\bibinfo{person}{Marcus Brandenburger},
  \bibinfo{person}{Christian Cachin}, \bibinfo{person}{Matthias Lorenz}, {and}
  \bibinfo{person}{R{\"{u}}diger Kapitza}.} \bibinfo{year}{2017}\natexlab{}.
\newblock \showarticletitle{Rollback and Forking Detection for Trusted
  Execution Environments Using Lightweight Collective Memory}. In
  \bibinfo{booktitle}{\emph{47th Annual {IEEE/IFIP} International Conference on
  Dependable Systems and Networks, {DSN} 2017, Denver, CO, USA, June 26-29,
  2017}}. \bibinfo{publisher}{IEEE}, \bibinfo{address}{Denver, CO, USA},
  \bibinfo{pages}{157--168}.
\newblock
\urldef\tempurl%
\url{https://doi.org/10.1109/DSN.2017.45}
\showDOI{\tempurl}


\bibitem[Bulck et~al\mbox{.}(2018)]%
        {van2018foreshadow}
\bibfield{author}{\bibinfo{person}{Jo~Van Bulck}, \bibinfo{person}{Marina
  Minkin}, \bibinfo{person}{Ofir Weisse}, \bibinfo{person}{Daniel Genkin},
  \bibinfo{person}{Baris Kasikci}, \bibinfo{person}{Frank Piessens},
  \bibinfo{person}{Mark Silberstein}, \bibinfo{person}{Thomas~F. Wenisch},
  \bibinfo{person}{Yuval Yarom}, {and} \bibinfo{person}{Raoul Strackx}.}
  \bibinfo{year}{2018}\natexlab{}.
\newblock \showarticletitle{Foreshadow: Extracting the Keys to the Intel {SGX}
  Kingdom with Transient {Out-of-Order} Execution}. In
  \bibinfo{booktitle}{\emph{27th USENIX Security Symposium (USENIX Security
  18)}}. \bibinfo{publisher}{USENIX Association}, \bibinfo{address}{Baltimore,
  MD}, \bibinfo{pages}{991{\textendash}1008}.
\newblock
\showISBNx{978-1-939133-04-5}
\urldef\tempurl%
\url{https://www.usenix.org/conference/usenixsecurity18/presentation/bulck}
\showURL{%
\tempurl}


\bibitem[Castro(2001)]%
        {castrothesis}
\bibfield{author}{\bibinfo{person}{Miguel Castro}.}
  \bibinfo{year}{2001}\natexlab{}.
\newblock \emph{\bibinfo{title}{Practical Byzantine Fault Tolerance}}.
\newblock {Ph.D.} \bibinfo{school}{MIT}.
\newblock
\newblock
\shownote{Also as Technical Report MIT-LCS-TR-817}.


\bibitem[Castro and Liskov(1999)]%
        {castro1999practical}
\bibfield{author}{\bibinfo{person}{Miguel Castro} {and}
  \bibinfo{person}{Barbara Liskov}.} \bibinfo{year}{1999}\natexlab{}.
\newblock \showarticletitle{Practical Byzantine Fault Tolerance}. In
  \bibinfo{booktitle}{\emph{Proceedings of the Third Symposium on Operating
  Systems Design and Implementation}} (New Orleans, Louisiana, USA)
  \emph{(\bibinfo{series}{OSDI '99})}. \bibinfo{publisher}{USENIX Association},
  \bibinfo{address}{USA}, \bibinfo{pages}{173--186}.
\newblock
\showISBNx{1880446391}
\urldef\tempurl%
\url{https://dl.acm.org/doi/10.5555/296806.296824}
\showURL{%
\tempurl}


\bibitem[Chen et~al\mbox{.}(2021)]%
        {263816}
\bibfield{author}{\bibinfo{person}{Zitai Chen}, \bibinfo{person}{Georgios
  Vasilakis}, \bibinfo{person}{Kit Murdock}, \bibinfo{person}{Edward Dean},
  \bibinfo{person}{David Oswald}, {and} \bibinfo{person}{Flavio~D. Garcia}.}
  \bibinfo{year}{2021}\natexlab{}.
\newblock \showarticletitle{VoltPillager: Hardware-based fault injection
  attacks against Intel SGX Enclaves using the {SVID} voltage scaling
  interface}. In \bibinfo{booktitle}{\emph{30th USENIX Security Symposium
  (USENIX Security 21)}}. \bibinfo{publisher}{USENIX Association},
  \bibinfo{pages}{699--716}.
\newblock
\showISBNx{978-1-939133-24-3}
\urldef\tempurl%
\url{https://www.usenix.org/conference/usenixsecurity21/presentation/chen-zitai}
\showURL{%
\tempurl}


\bibitem[Cheng et~al\mbox{.}(2019)]%
        {cheng2019ekiden}
\bibfield{author}{\bibinfo{person}{Raymond Cheng}, \bibinfo{person}{Fan Zhang},
  \bibinfo{person}{Jernej Kos}, \bibinfo{person}{Warren He},
  \bibinfo{person}{Nicholas Hynes}, \bibinfo{person}{Noah Johnson},
  \bibinfo{person}{Ari Juels}, \bibinfo{person}{Andrew Miller}, {and}
  \bibinfo{person}{Dawn Song}.} \bibinfo{year}{2019}\natexlab{}.
\newblock \showarticletitle{Ekiden: A Platform for Confidentiality-Preserving,
  Trustworthy, and Performant Smart Contracts}. In
  \bibinfo{booktitle}{\emph{2019 IEEE European Symposium on Security and
  Privacy (EuroS P)}}. \bibinfo{publisher}{IEEE}, \bibinfo{address}{Stockholm,
  Sweden}, \bibinfo{pages}{185--200}.
\newblock
\urldef\tempurl%
\url{https://doi.org/10.1109/EuroSP.2019.00023}
\showDOI{\tempurl}


\bibitem[Clement et~al\mbox{.}(2012)]%
        {clement2012limited}
\bibfield{author}{\bibinfo{person}{Allen Clement}, \bibinfo{person}{Flavio
  Junqueira}, \bibinfo{person}{Aniket Kate}, {and} \bibinfo{person}{Rodrigo
  Rodrigues}.} \bibinfo{year}{2012}\natexlab{}.
\newblock \showarticletitle{On the (Limited) Power of Non-Equivocation}
  \emph{(\bibinfo{series}{PODC '12})}. \bibinfo{publisher}{ACM},
  \bibinfo{address}{New York, NY, USA}, \bibinfo{pages}{301--308}.
\newblock
\showISBNx{9781450314503}
\urldef\tempurl%
\url{https://doi.org/10.1145/2332432.2332490}
\showDOI{\tempurl}


\bibitem[Clement et~al\mbox{.}(2009a)]%
        {clement09sosp}
\bibfield{author}{\bibinfo{person}{Allen Clement}, \bibinfo{person}{Manos
  Kapritsos}, \bibinfo{person}{Sangmin Lee}, \bibinfo{person}{Yang Wang},
  \bibinfo{person}{Lorenzo Alvisi}, \bibinfo{person}{Mike Dahlin}, {and}
  \bibinfo{person}{Taylor Riche}.} \bibinfo{year}{2009}\natexlab{a}.
\newblock \showarticletitle{Upright Cluster Services}. In
  \bibinfo{booktitle}{\emph{Proceedings of the ACM SIGOPS 22nd Symposium on
  Operating Systems Principles}} (Big Sky, Montana, USA)
  \emph{(\bibinfo{series}{SOSP '09})}. \bibinfo{publisher}{ACM},
  \bibinfo{address}{New York, NY, USA}, \bibinfo{pages}{277--290}.
\newblock
\showISBNx{9781605587523}
\urldef\tempurl%
\url{https://doi.org/10.1145/1629575.1629602}
\showDOI{\tempurl}


\bibitem[Clement et~al\mbox{.}(2009b)]%
        {clement2009making}
\bibfield{author}{\bibinfo{person}{Allen Clement}, \bibinfo{person}{Edmund
  Wong}, \bibinfo{person}{Lorenzo Alvisi}, \bibinfo{person}{Mike Dahlin}, {and}
  \bibinfo{person}{Mirco Marchetti}.} \bibinfo{year}{2009}\natexlab{b}.
\newblock \showarticletitle{Making Byzantine Fault Tolerant Systems Tolerate
  Byzantine Faults}. In \bibinfo{booktitle}{\emph{Proceedings of the 6th USENIX
  Symposium on Networked Systems Design and Implementation}} (Boston,
  Massachusetts) \emph{(\bibinfo{series}{NSDI'09})}. \bibinfo{publisher}{USENIX
  Association}, \bibinfo{address}{USA}, \bibinfo{pages}{153--168}.
\newblock


\bibitem[Cloosters et~al\mbox{.}(2020)]%
        {teerex}
\bibfield{author}{\bibinfo{person}{Tobias Cloosters}, \bibinfo{person}{Michael
  Rodler}, {and} \bibinfo{person}{Lucas Davi}.}
  \bibinfo{year}{2020}\natexlab{}.
\newblock \showarticletitle{TeeRex: Discovery and Exploitation of Memory
  Corruption Vulnerabilities in SGX Enclaves}. In
  \bibinfo{booktitle}{\emph{29th USENIX Security Symposium (USENIX Security
  20)}}. \bibinfo{publisher}{USENIX Association}, \bibinfo{address}{Boston, MA,
  USA}, \bibinfo{pages}{841--858}.
\newblock
\showISBNx{978-1-939133-17-5}
\urldef\tempurl%
\url{https://www.usenix.org/conference/usenixsecurity20/presentation/cloosters}
\showURL{%
\tempurl}


\bibitem[Corp(2018)]%
        {retail}
\bibfield{author}{\bibinfo{person}{IBM Corp}.} \bibinfo{year}{2018}\natexlab{}.
\newblock \bibinfo{booktitle}{\emph{Blockchain in retail solutions}}.
\newblock
\urldef\tempurl%
\url{https://www.ibm.com/blockchain/industries/retail}
\showURL{%
Retrieved July 7, 2021 from \tempurl}


\bibitem[Corporation(2019)]%
        {confidentialc}
\bibfield{author}{\bibinfo{person}{Intel Corporation}.}
  \bibinfo{year}{2019}\natexlab{}.
\newblock \bibinfo{booktitle}{\emph{Confidential Computing Consortium}}.
\newblock
\urldef\tempurl%
\url{https://www.intel.com/content/www/us/en/security/confidential-computing.html}
\showURL{%
Retrieved October 2, 2021 from \tempurl}


\bibitem[Corporation(2021)]%
        {usecases}
\bibfield{author}{\bibinfo{person}{IBM Corporation}.}
  \bibinfo{year}{2021}\natexlab{}.
\newblock \bibinfo{booktitle}{\emph{Research leading blockchain use cases}}.
\newblock
\urldef\tempurl%
\url{https://www.ibm.com/blockchain/use-cases/}
\showURL{%
Retrieved September 16, 2021 from \tempurl}


\bibitem[Costan and Devadas(2016)]%
        {costan2016intel}
\bibfield{author}{\bibinfo{person}{Victor Costan} {and}
  \bibinfo{person}{Srinivas Devadas}.} \bibinfo{year}{2016}\natexlab{}.
\newblock \showarticletitle{Intel sgx explained.}
\newblock \bibinfo{journal}{\emph{IACR Cryptol. ePrint Arch.}}
  \bibinfo{volume}{2016}, \bibinfo{number}{86} (\bibinfo{year}{2016}),
  \bibinfo{pages}{1--118}.
\newblock


\bibitem[Decouchant et~al\mbox{.}(2022)]%
        {decouchant2022damysus}
\bibfield{author}{\bibinfo{person}{J\'{e}r\'{e}mie Decouchant},
  \bibinfo{person}{David Kozhaya}, \bibinfo{person}{Vincent Rahli}, {and}
  \bibinfo{person}{Jiangshan Yu}.} \bibinfo{year}{2022}\natexlab{}.
\newblock \showarticletitle{DAMYSUS: Streamlined BFT Consensus Leveraging
  Trusted Components}. In \bibinfo{booktitle}{\emph{Proceedings of the
  Seventeenth European Conference on Computer Systems}} (Rennes, France)
  \emph{(\bibinfo{series}{EuroSys '22})}. \bibinfo{publisher}{ACM},
  \bibinfo{address}{New York, NY, USA}, \bibinfo{pages}{1--16}.
\newblock
\showISBNx{9781450391627}
\urldef\tempurl%
\url{https://doi.org/10.1145/3492321.3519568}
\showDOI{\tempurl}


\bibitem[Dettoni et~al\mbox{.}(2013)]%
        {dettoni2013byzantine}
\bibfield{author}{\bibinfo{person}{Fernando Dettoni},
  \bibinfo{person}{Lau~Cheuk Lung}, \bibinfo{person}{Miguel Correia}, {and}
  \bibinfo{person}{Aldelir~Fernando Luiz}.} \bibinfo{year}{2013}\natexlab{}.
\newblock \showarticletitle{Byzantine fault-tolerant state machine replication
  with twin virtual machines}. In \bibinfo{booktitle}{\emph{2013 IEEE Symposium
  on Computers and Communications (ISCC)}}. \bibinfo{publisher}{IEEE},
  \bibinfo{address}{Split, Croatia}, \bibinfo{pages}{398--403}.
\newblock
\urldef\tempurl%
\url{https://doi.org/10.1109/ISCC.2013.6754979}
\showDOI{\tempurl}


\bibitem[Didona and Zwaenepoel(2019)]%
        {didona2019size}
\bibfield{author}{\bibinfo{person}{Diego Didona} {and} \bibinfo{person}{Willy
  Zwaenepoel}.} \bibinfo{year}{2019}\natexlab{}.
\newblock \showarticletitle{Size-Aware Sharding for Improving Tail Latencies in
  in-Memory Key-Value Stores}. In \bibinfo{booktitle}{\emph{Proceedings of the
  16th USENIX Conference on Networked Systems Design and Implementation}}
  (Boston, MA, USA) \emph{(\bibinfo{series}{NSDI'19})}.
  \bibinfo{publisher}{USENIX Association}, \bibinfo{address}{USA},
  \bibinfo{pages}{79--93}.
\newblock
\showISBNx{9781931971492}


\bibitem[Distler et~al\mbox{.}(2011)]%
        {distler2011spare}
\bibfield{author}{\bibinfo{person}{Tobias Distler}, \bibinfo{person}{Ivan
  Popov}, \bibinfo{person}{Wolfgang Schr{\"o}der-Preikschat},
  \bibinfo{person}{Hans~P Reiser}, {and} \bibinfo{person}{R{\"u}diger
  Kapitza}.} \bibinfo{year}{2011}\natexlab{}.
\newblock \showarticletitle{SPARE: Replicas on Hold}. In
  \bibinfo{booktitle}{\emph{Proceedings of the 18th Network and Distributed
  System Security Symposium (NDSS '11)}}. \bibinfo{publisher}{Internet
  Society}, \bibinfo{address}{San Diego, California, USA},
  \bibinfo{pages}{407--420}.
\newblock


\bibitem[Garcia et~al\mbox{.}(2019)]%
        {garcia2019lazarus}
\bibfield{author}{\bibinfo{person}{Miguel Garcia}, \bibinfo{person}{Alysson
  Bessani}, {and} \bibinfo{person}{Nuno Neves}.}
  \bibinfo{year}{2019}\natexlab{}.
\newblock \showarticletitle{Lazarus: Automatic Management of Diversity in BFT
  Systems}. In \bibinfo{booktitle}{\emph{Proceedings of the 20th International
  Middleware Conference}} (Davis, CA, USA) \emph{(\bibinfo{series}{Middleware
  '19})}. \bibinfo{publisher}{ACM}, \bibinfo{address}{New York, NY, USA},
  \bibinfo{pages}{241--254}.
\newblock
\showISBNx{9781450370097}
\urldef\tempurl%
\url{https://doi.org/10.1145/3361525.3361550}
\showDOI{\tempurl}


\bibitem[Golan~Gueta et~al\mbox{.}(2019)]%
        {sbft}
\bibfield{author}{\bibinfo{person}{Guy Golan~Gueta}, \bibinfo{person}{Ittai
  Abraham}, \bibinfo{person}{Shelly Grossman}, \bibinfo{person}{Dahlia Malkhi},
  \bibinfo{person}{Benny Pinkas}, \bibinfo{person}{Michael Reiter},
  \bibinfo{person}{Dragos-Adrian Seredinschi}, \bibinfo{person}{Orr Tamir},
  {and} \bibinfo{person}{Alin Tomescu}.} \bibinfo{year}{2019}\natexlab{}.
\newblock \showarticletitle{SBFT: A Scalable and Decentralized Trust
  Infrastructure}. In \bibinfo{booktitle}{\emph{2019 49th Annual IEEE/IFIP
  International Conference on Dependable Systems and Networks (DSN)}}.
  \bibinfo{publisher}{IEEE}, \bibinfo{address}{Portland, OR, USA},
  \bibinfo{pages}{568--580}.
\newblock
\urldef\tempurl%
\url{https://doi.org/10.1109/DSN.2019.00063}
\showDOI{\tempurl}


\bibitem[Guo et~al\mbox{.}(2020)]%
        {dumbo}
\bibfield{author}{\bibinfo{person}{Bingyong Guo}, \bibinfo{person}{Zhenliang
  Lu}, \bibinfo{person}{Qiang Tang}, \bibinfo{person}{Jing Xu}, {and}
  \bibinfo{person}{Zhenfeng Zhang}.} \bibinfo{year}{2020}\natexlab{}.
\newblock \bibinfo{booktitle}{\emph{Dumbo: Faster Asynchronous BFT Protocols}}.
\newblock \bibinfo{publisher}{ACM}, \bibinfo{address}{New York, NY, USA},
  \bibinfo{pages}{803--818}.
\newblock
\showISBNx{9781450370899}
\urldef\tempurl%
\url{https://doi.org/10.1145/3372297.3417262}
\showURL{%
\tempurl}


\bibitem[Inc(2022)]%
        {teaclave}
\bibfield{author}{\bibinfo{person}{Apache Inc}.}
  \bibinfo{year}{2022}\natexlab{}.
\newblock \bibinfo{title}{Teaclave SGX SDK}.
\newblock
\newblock
\urldef\tempurl%
\url{https://github.com/apache/incubator-teaclave-sgx-sdk}
\showURL{%
\tempurl}


\bibitem[Kapitza et~al\mbox{.}(2012)]%
        {cheapbft}
\bibfield{author}{\bibinfo{person}{R\"{u}diger Kapitza},
  \bibinfo{person}{Johannes Behl}, \bibinfo{person}{Christian Cachin},
  \bibinfo{person}{Tobias Distler}, \bibinfo{person}{Simon Kuhnle},
  \bibinfo{person}{Seyed~Vahid Mohammadi}, \bibinfo{person}{Wolfgang
  Schr\"{o}der-Preikschat}, {and} \bibinfo{person}{Klaus Stengel}.}
  \bibinfo{year}{2012}\natexlab{}.
\newblock \showarticletitle{CheapBFT: Resource-Efficient Byzantine Fault
  Tolerance}. In \bibinfo{booktitle}{\emph{Proceedings of the 7th ACM European
  Conference on Computer Systems}} (Bern, Switzerland)
  \emph{(\bibinfo{series}{EuroSys '12})}. \bibinfo{publisher}{ACM},
  \bibinfo{address}{New York, NY, USA}, \bibinfo{pages}{295--308}.
\newblock
\showISBNx{9781450312233}
\urldef\tempurl%
\url{https://doi.org/10.1145/2168836.2168866}
\showDOI{\tempurl}


\bibitem[Kim et~al\mbox{.}(2019)]%
        {kim2019shieldstore}
\bibfield{author}{\bibinfo{person}{Taehoon Kim}, \bibinfo{person}{Joongun
  Park}, \bibinfo{person}{Jaewook Woo}, \bibinfo{person}{Seungheun Jeon}, {and}
  \bibinfo{person}{Jaehyuk Huh}.} \bibinfo{year}{2019}\natexlab{}.
\newblock \showarticletitle{ShieldStore: Shielded In-Memory Key-Value Storage
  with SGX}. In \bibinfo{booktitle}{\emph{Proceedings of the Fourteenth EuroSys
  Conference 2019}} (Dresden, Germany) \emph{(\bibinfo{series}{EuroSys '19})}.
  \bibinfo{publisher}{ACM}, \bibinfo{address}{New York, NY, USA}, Article
  \bibinfo{articleno}{14}, \bibinfo{numpages}{15}~pages.
\newblock
\showISBNx{9781450362818}
\urldef\tempurl%
\url{https://doi.org/10.1145/3302424.3303951}
\showDOI{\tempurl}


\bibitem[Kocher et~al\mbox{.}(2019)]%
        {kocher2019spectre}
\bibfield{author}{\bibinfo{person}{Paul Kocher}, \bibinfo{person}{Jann Horn},
  \bibinfo{person}{Anders Fogh}, \bibinfo{person}{Daniel Genkin},
  \bibinfo{person}{Daniel Gruss}, \bibinfo{person}{Werner Haas},
  \bibinfo{person}{Mike Hamburg}, \bibinfo{person}{Moritz Lipp},
  \bibinfo{person}{Stefan Mangard}, \bibinfo{person}{Thomas Prescher},
  \bibinfo{person}{Michael Schwarz}, {and} \bibinfo{person}{Yuval Yarom}.}
  \bibinfo{year}{2019}\natexlab{}.
\newblock \showarticletitle{Spectre Attacks: Exploiting Speculative Execution}.
  In \bibinfo{booktitle}{\emph{2019 IEEE Symposium on Security and Privacy
  (SP)}}. \bibinfo{publisher}{IEEE}, \bibinfo{address}{San Francisco, CA, USA},
  \bibinfo{pages}{1--19}.
\newblock
\urldef\tempurl%
\url{https://doi.org/10.1109/SP.2019.00002}
\showDOI{\tempurl}


\bibitem[Kotla et~al\mbox{.}(2010)]%
        {kotla2010zyzzyva}
\bibfield{author}{\bibinfo{person}{Ramakrishna Kotla}, \bibinfo{person}{Lorenzo
  Alvisi}, \bibinfo{person}{Mike Dahlin}, \bibinfo{person}{Allen Clement},
  {and} \bibinfo{person}{Edmund Wong}.} \bibinfo{year}{2010}\natexlab{}.
\newblock \showarticletitle{Zyzzyva: Speculative Byzantine Fault Tolerance}.
\newblock \bibinfo{journal}{\emph{ACM Trans. Comput. Syst.}}
  \bibinfo{volume}{27}, \bibinfo{number}{4}, Article \bibinfo{articleno}{7}
  (\bibinfo{date}{jan} \bibinfo{year}{2010}), \bibinfo{numpages}{39}~pages.
\newblock
\showISSN{0734-2071}
\urldef\tempurl%
\url{https://doi.org/10.1145/1658357.1658358}
\showDOI{\tempurl}


\bibitem[Kuo et~al\mbox{.}(2017)]%
        {kuo2017blockchain}
\bibfield{author}{\bibinfo{person}{Tsung-Ting Kuo}, \bibinfo{person}{Hyeon-Eui
  Kim}, {and} \bibinfo{person}{Lucila Ohno-Machado}.}
  \bibinfo{year}{2017}\natexlab{}.
\newblock \showarticletitle{Blockchain distributed ledger technologies for
  biomedical and health care applications}.
\newblock \bibinfo{journal}{\emph{Journal of the American Medical Informatics
  Association}} \bibinfo{volume}{24}, \bibinfo{number}{6}
  (\bibinfo{year}{2017}), \bibinfo{pages}{1211--1220}.
\newblock


\bibitem[Lamport et~al\mbox{.}(1982)]%
        {lamport2019byzantine}
\bibfield{author}{\bibinfo{person}{Leslie Lamport}, \bibinfo{person}{Robert
  Shostak}, {and} \bibinfo{person}{Marshall Pease}.}
  \bibinfo{year}{1982}\natexlab{}.
\newblock \showarticletitle{The Byzantine Generals Problem}.
\newblock \bibinfo{journal}{\emph{ACM Trans. Program. Lang. Syst.}}
  \bibinfo{volume}{4}, \bibinfo{number}{3} (\bibinfo{date}{jul}
  \bibinfo{year}{1982}), \bibinfo{pages}{382--401}.
\newblock
\showISSN{0164-0925}
\urldef\tempurl%
\url{https://doi.org/10.1145/357172.357176}
\showDOI{\tempurl}


\bibitem[Li et~al\mbox{.}(2018)]%
        {troxy}
\bibfield{author}{\bibinfo{person}{Bijun Li}, \bibinfo{person}{Nico
  Weichbrodt}, \bibinfo{person}{Johannes Behl}, \bibinfo{person}{Pierre-Louis
  Aublin}, \bibinfo{person}{Tobias Distler}, {and} \bibinfo{person}{R\"{u}diger
  Kapitza}.} \bibinfo{year}{2018}\natexlab{}.
\newblock \showarticletitle{Troxy: Transparent Access to Byzantine
  Fault-Tolerant Systems}. In \bibinfo{booktitle}{\emph{2018 48th Annual
  IEEE/IFIP International Conference on Dependable Systems and Networks
  (DSN)}}. \bibinfo{publisher}{IEEE}, \bibinfo{address}{Luxembourg,
  Luxembourg}, \bibinfo{pages}{59--70}.
\newblock
\showISSN{2158-3927}
\urldef\tempurl%
\url{https://doi.org/10.1109/DSN.2018.00019}
\showDOI{\tempurl}


\bibitem[Li and Mazi\'{e}res(2007)]%
        {li2007beyond}
\bibfield{author}{\bibinfo{person}{Jinyuan Li} {and} \bibinfo{person}{David
  Mazi\'{e}res}.} \bibinfo{year}{2007}\natexlab{}.
\newblock \showarticletitle{Beyond One-Third Faulty Replicas in Byzantine Fault
  Tolerant Systems}. In \bibinfo{booktitle}{\emph{Proceedings of the 4th USENIX
  Conference on Networked Systems Design Implementation}} (Cambridge, MA)
  \emph{(\bibinfo{series}{NSDI'07})}. \bibinfo{publisher}{USENIX Association},
  \bibinfo{address}{USA}, \bibinfo{pages}{10}.
\newblock


\bibitem[Liu et~al\mbox{.}(2019)]%
        {liu2018scalable}
\bibfield{author}{\bibinfo{person}{Jian Liu}, \bibinfo{person}{Wenting Li},
  \bibinfo{person}{Ghassan~O. Karame}, {and} \bibinfo{person}{N. Asokan}.}
  \bibinfo{year}{2019}\natexlab{}.
\newblock \showarticletitle{Scalable Byzantine Consensus via Hardware-Assisted
  Secret Sharing}.
\newblock \bibinfo{journal}{\emph{IEEE Trans. Comput.}} \bibinfo{volume}{68},
  \bibinfo{number}{1} (\bibinfo{year}{2019}), \bibinfo{pages}{139--151}.
\newblock
\urldef\tempurl%
\url{https://doi.org/10.1109/TC.2018.2860009}
\showDOI{\tempurl}


\bibitem[Liu et~al\mbox{.}(2016)]%
        {liu16osdi}
\bibfield{author}{\bibinfo{person}{Shengyun Liu}, \bibinfo{person}{Paolo
  Viotti}, \bibinfo{person}{Christian Cachin}, \bibinfo{person}{Vivien Quema},
  {and} \bibinfo{person}{Marko Vukoli\'{c}}.} \bibinfo{year}{2016}\natexlab{}.
\newblock \showarticletitle{XFT: Practical Fault Tolerance beyond Crashes}. In
  \bibinfo{booktitle}{\emph{12th USENIX Symposium on Operating Systems Design
  and Implementation (OSDI 16)}}. \bibinfo{publisher}{USENIX Association},
  \bibinfo{address}{Savannah, GA}, \bibinfo{pages}{485--500}.
\newblock
\showISBNx{978-1-931971-33-1}
\urldef\tempurl%
\url{https://www.usenix.org/conference/osdi16/technical-sessions/presentation/liu}
\showURL{%
\tempurl}


\bibitem[Malkhi et~al\mbox{.}(2019)]%
        {malkhi2019flexible}
\bibfield{author}{\bibinfo{person}{Dahlia Malkhi}, \bibinfo{person}{Kartik
  Nayak}, {and} \bibinfo{person}{Ling Ren}.} \bibinfo{year}{2019}\natexlab{}.
\newblock \showarticletitle{Flexible Byzantine Fault Tolerance}. In
  \bibinfo{booktitle}{\emph{Proceedings of the 2019 ACM SIGSAC Conference on
  Computer and Communications Security}} (London, United Kingdom)
  \emph{(\bibinfo{series}{CCS '19})}. \bibinfo{publisher}{ACM},
  \bibinfo{address}{New York, NY, USA}, \bibinfo{pages}{1041--1053}.
\newblock
\showISBNx{9781450367479}
\urldef\tempurl%
\url{https://doi.org/10.1145/3319535.3354225}
\showDOI{\tempurl}


\bibitem[Martin and Alvisi(2005)]%
        {martin2006fast}
\bibfield{author}{\bibinfo{person}{J.-P. Martin} {and} \bibinfo{person}{L.
  Alvisi}.} \bibinfo{year}{2005}\natexlab{}.
\newblock \showarticletitle{Fast Byzantine consensus}. In
  \bibinfo{booktitle}{\emph{2005 International Conference on Dependable Systems
  and Networks (DSN'05)}}. \bibinfo{publisher}{IEEE},
  \bibinfo{address}{Yokohama, Japan}, \bibinfo{pages}{402--411}.
\newblock
\urldef\tempurl%
\url{https://doi.org/10.1109/DSN.2005.48}
\showDOI{\tempurl}


\bibitem[McMillan and Padon(2020)]%
        {ivy}
\bibfield{author}{\bibinfo{person}{Kenneth~L. McMillan} {and}
  \bibinfo{person}{Oded Padon}.} \bibinfo{year}{2020}\natexlab{}.
\newblock \showarticletitle{Ivy: A Multi-Modal Verification Tool for
  Distributed Algorithms}. In \bibinfo{booktitle}{\emph{Computer Aided
  Verification: 32nd International Conference, CAV 2020, Los Angeles, CA, USA,
  July 21--24, 2020, Proceedings, Part II}} (Los Angeles, CA, USA).
  \bibinfo{publisher}{Springer-Verlag}, \bibinfo{address}{Berlin, Heidelberg},
  \bibinfo{pages}{190--202}.
\newblock
\showISBNx{978-3-030-53290-1}
\urldef\tempurl%
\url{https://doi.org/10.1007/978-3-030-53291-8_12}
\showDOI{\tempurl}


\bibitem[Oleksenko et~al\mbox{.}(2018)]%
        {oleksenko2018varys}
\bibfield{author}{\bibinfo{person}{Oleksii Oleksenko}, \bibinfo{person}{Bohdan
  Trach}, \bibinfo{person}{Robert Krahn}, \bibinfo{person}{Andre Martin},
  \bibinfo{person}{Christof Fetzer}, {and} \bibinfo{person}{Mark Silberstein}.}
  \bibinfo{year}{2018}\natexlab{}.
\newblock \showarticletitle{Varys: Protecting SGX Enclaves from Practical
  Side-Channel Attacks}. In \bibinfo{booktitle}{\emph{Proceedings of the 2018
  USENIX Conference on Usenix Annual Technical Conference}} (Boston, MA, USA)
  \emph{(\bibinfo{series}{USENIX ATC '18})}. \bibinfo{publisher}{USENIX
  Association}, \bibinfo{address}{USA}, \bibinfo{pages}{227--239}.
\newblock
\showISBNx{9781931971447}


\bibitem[Porto et~al\mbox{.}(2015)]%
        {porto15eurosys}
\bibfield{author}{\bibinfo{person}{Daniel Porto}, \bibinfo{person}{Jo\~{a}o
  Leit\~{a}o}, \bibinfo{person}{Cheng Li}, \bibinfo{person}{Allen Clement},
  \bibinfo{person}{Aniket Kate}, \bibinfo{person}{Flavio Junqueira}, {and}
  \bibinfo{person}{Rodrigo Rodrigues}.} \bibinfo{year}{2015}\natexlab{}.
\newblock \showarticletitle{Visigoth Fault Tolerance}. In
  \bibinfo{booktitle}{\emph{Proceedings of the Tenth European Conference on
  Computer Systems}} (Bordeaux, France) \emph{(\bibinfo{series}{EuroSys '15})}.
  \bibinfo{publisher}{ACM}, \bibinfo{address}{New York, NY, USA}, Article
  \bibinfo{articleno}{8}, \bibinfo{numpages}{14}~pages.
\newblock
\showISBNx{9781450332385}
\urldef\tempurl%
\url{https://doi.org/10.1145/2741948.2741979}
\showDOI{\tempurl}


\bibitem[Puddu et~al\mbox{.}(2021)]%
        {274699}
\bibfield{author}{\bibinfo{person}{Ivan Puddu}, \bibinfo{person}{Moritz
  Schneider}, \bibinfo{person}{Miro Haller}, {and} \bibinfo{person}{Srdjan
  Capkun}.} \bibinfo{year}{2021}\natexlab{}.
\newblock \showarticletitle{Frontal Attack: Leaking Control-Flow in SGX via the
  {CPU} Frontend}. In \bibinfo{booktitle}{\emph{30th USENIX Security Symposium
  (USENIX Security 21)}}. \bibinfo{publisher}{USENIX Association},
  \bibinfo{pages}{663--680}.
\newblock
\showISBNx{978-1-939133-24-3}
\urldef\tempurl%
\url{https://www.usenix.org/conference/usenixsecurity21/presentation/puddu}
\showURL{%
\tempurl}


\bibitem[R\"{u}sch et~al\mbox{.}(2019)]%
        {rusch2019themis}
\bibfield{author}{\bibinfo{person}{Signe R\"{u}sch}, \bibinfo{person}{Kai
  Bleeke}, {and} \bibinfo{person}{R\"{u}diger Kapitza}.}
  \bibinfo{year}{2019}\natexlab{}.
\newblock \showarticletitle{Themis: An Efficient and Memory-Safe BFT Framework
  in Rust: Research Statement}. In \bibinfo{booktitle}{\emph{Proceedings of the
  3rd Workshop on Scalable and Resilient Infrastructures for Distributed
  Ledgers}} (Davis, CA, USA) \emph{(\bibinfo{series}{SERIAL '19})}.
  \bibinfo{publisher}{ACM}, \bibinfo{address}{New York, NY, USA},
  \bibinfo{pages}{9--10}.
\newblock
\showISBNx{9781450370295}
\urldef\tempurl%
\url{https://doi.org/10.1145/3366611.3368144}
\showDOI{\tempurl}


\bibitem[Russinovich et~al\mbox{.}(2019)]%
        {russinovich2019ccf}
\bibfield{author}{\bibinfo{person}{Mark Russinovich}, \bibinfo{person}{Edward
  Ashton}, \bibinfo{person}{Christine Avanessians}, \bibinfo{person}{Miguel
  Castro}, \bibinfo{person}{Amaury Chamayou}, \bibinfo{person}{Sylvan Clebsch},
  \bibinfo{person}{Manuel Costa}, \bibinfo{person}{C\'{e}dric Fournet},
  \bibinfo{person}{Matthew Kerner}, \bibinfo{person}{Sid Krishna},
  \bibinfo{person}{Julien Maffre}, \bibinfo{person}{Thomas Moscibroda},
  \bibinfo{person}{Kartik Nayak}, \bibinfo{person}{Olya Ohrimenko},
  \bibinfo{person}{Felix Schuster}, \bibinfo{person}{Roy Schwartz},
  \bibinfo{person}{Alex Shamis}, \bibinfo{person}{Olga Vrousgou}, {and}
  \bibinfo{person}{Christoph~M. Wintersteiger}.}
  \bibinfo{year}{2019}\natexlab{}.
\newblock \bibinfo{booktitle}{\emph{CCF: A Framework for Building Confidential
  Verifiable Replicated Services}}.
\newblock \bibinfo{type}{{T}echnical {R}eport} MSR-TR-2019-16.
  \bibinfo{institution}{Microsoft}.
\newblock
\urldef\tempurl%
\url{https://www.microsoft.com/en-us/research/publication/ccf-a-framework-for-building-confidential-verifiable-replicated-services/}
\showURL{%
\tempurl}


\bibitem[Seo et~al\mbox{.}(2017)]%
        {seo2017sgx}
\bibfield{author}{\bibinfo{person}{Jaebaek Seo}, \bibinfo{person}{Byoungyoung
  Lee}, \bibinfo{person}{Seong~Min Kim}, \bibinfo{person}{Ming-Wei Shih},
  \bibinfo{person}{Insik Shin}, \bibinfo{person}{Dongsu Han}, {and}
  \bibinfo{person}{Taesoo Kim}.} \bibinfo{year}{2017}\natexlab{}.
\newblock \showarticletitle{SGX-Shield: Enabling Address Space Layout
  Randomization for SGX Programs}. In \bibinfo{booktitle}{\emph{NDSS}}.
  \bibinfo{publisher}{Internet Society}, \bibinfo{address}{San Diego, CA USA}.
\newblock


\bibitem[Shih et~al\mbox{.}(2017)]%
        {shih2017t}
\bibfield{author}{\bibinfo{person}{Ming-Wei Shih}, \bibinfo{person}{Sangho
  Lee}, \bibinfo{person}{Taesoo Kim}, {and} \bibinfo{person}{Marcus Peinado}.}
  \bibinfo{year}{2017}\natexlab{}.
\newblock \showarticletitle{T-SGX: Eradicating Controlled-Channel Attacks
  Against Enclave Programs}. In \bibinfo{booktitle}{\emph{Network and
  Distributed System Security Symposium 2017 (NDSS'17)}
  (\bibinfo{edition}{network and distributed system security symposium 2017
  (ndss'17)} ed.)}. \bibinfo{publisher}{Internet Society},
  \bibinfo{address}{San Diego, CA USA}.
\newblock
\urldef\tempurl%
\url{https://www.microsoft.com/en-us/research/publication/t-sgx-eradicating-controlled-channel-attacks-enclave-programs/}
\showURL{%
\tempurl}


\bibitem[Smith(2022)]%
        {ring}
\bibfield{author}{\bibinfo{person}{Brian Smith}.}
  \bibinfo{year}{2022}\natexlab{}.
\newblock \bibinfo{title}{ring}.
\newblock
\newblock
\urldef\tempurl%
\url{https://github.com/briansmith/ring}
\showURL{%
\tempurl}


\bibitem[Stathakopoulou et~al\mbox{.}(2021)]%
        {stathakopoulou2021adding}
\bibfield{author}{\bibinfo{person}{Chrysoula Stathakopoulou},
  \bibinfo{person}{Signe R\"usch}, \bibinfo{person}{Marcus Brandenburger},
  {and} \bibinfo{person}{Marko Vukoli\'{c}}.} \bibinfo{year}{2021}\natexlab{}.
\newblock \showarticletitle{Adding Fairness to Order: Preventing Front-Running
  Attacks in BFT Protocols using TEEs}. In \bibinfo{booktitle}{\emph{2021 40th
  International Symposium on Reliable Distributed Systems (SRDS)}}.
  \bibinfo{publisher}{IEEE}, \bibinfo{address}{Chicago, IL, USA},
  \bibinfo{pages}{34--45}.
\newblock
\urldef\tempurl%
\url{https://doi.org/10.1109/SRDS53918.2021.00013}
\showDOI{\tempurl}


\bibitem[Taube et~al\mbox{.}(2018)]%
        {ivypbft}
\bibfield{author}{\bibinfo{person}{Marcelo Taube}, \bibinfo{person}{Giuliano
  Losa}, \bibinfo{person}{Kenneth~L. McMillan}, \bibinfo{person}{Oded Padon},
  \bibinfo{person}{Mooly Sagiv}, \bibinfo{person}{Sharon Shoham},
  \bibinfo{person}{James R.Wilcox}, {and} \bibinfo{person}{Doug Woos}.}
  \bibinfo{year}{2018}\natexlab{}.
\newblock \bibinfo{title}{Modularity for Decidability of Deductive Verification
  with Applications to Distributed Systems}.
\newblock
\newblock
\urldef\tempurl%
\url{https://doi.org/10.5281/zenodo.2577103}
\showURL{%
\tempurl}


\bibitem[Veronese et~al\mbox{.}(2011)]%
        {minbft}
\bibfield{author}{\bibinfo{person}{Giuliana~Santos Veronese},
  \bibinfo{person}{Miguel Correia}, \bibinfo{person}{Alysson~Neves Bessani},
  \bibinfo{person}{Lau~Cheuk Lung}, {and} \bibinfo{person}{Paulo Verissimo}.}
  \bibinfo{year}{2011}\natexlab{}.
\newblock \showarticletitle{Efficient byzantine fault-tolerance}.
\newblock \bibinfo{journal}{\emph{IEEE Trans. Comput.}} \bibinfo{volume}{62},
  \bibinfo{number}{1} (\bibinfo{year}{2011}), \bibinfo{pages}{16--30}.
\newblock


\bibitem[Wang et~al\mbox{.}(2022)]%
        {wang2022zero}
\bibfield{author}{\bibinfo{person}{Zhipeng Wang}, \bibinfo{person}{Stefanos
  Chaliasos}, \bibinfo{person}{Kaihua Qin}, \bibinfo{person}{Liyi Zhou},
  \bibinfo{person}{Lifeng Gao}, \bibinfo{person}{Pascal Berrang},
  \bibinfo{person}{Ben Livshits}, {and} \bibinfo{person}{Arthur Gervais}.}
  \bibinfo{year}{2022}\natexlab{}.
\newblock \bibinfo{title}{On How Zero-Knowledge Proof Blockchain Mixers
  Improve, and Worsen User Privacy}.
\newblock
\newblock
\urldef\tempurl%
\url{https://doi.org/10.48550/ARXIV.2201.09035}
\showDOI{\tempurl}


\bibitem[Weichbrodt et~al\mbox{.}(2016)]%
        {weichbrodt2016asyncshock}
\bibfield{author}{\bibinfo{person}{Nico Weichbrodt}, \bibinfo{person}{Anil
  Kurmus}, \bibinfo{person}{Peter Pietzuch}, {and} \bibinfo{person}{R{\"u}diger
  Kapitza}.} \bibinfo{year}{2016}\natexlab{}.
\newblock \showarticletitle{AsyncShock: Exploiting Synchronisation Bugs in
  Intel SGX Enclaves}. In \bibinfo{booktitle}{\emph{Computer Security --
  ESORICS 2016}}, \bibfield{editor}{\bibinfo{person}{Ioannis Askoxylakis},
  \bibinfo{person}{Sotiris Ioannidis}, \bibinfo{person}{Sokratis Katsikas},
  {and} \bibinfo{person}{Catherine Meadows}} (Eds.).
  \bibinfo{publisher}{Springer}, \bibinfo{address}{Cham},
  \bibinfo{pages}{440--457}.
\newblock
\showISBNx{978-3-319-45744-4}


\bibitem[Weisse et~al\mbox{.}(2017)]%
        {weisse2017regaining}
\bibfield{author}{\bibinfo{person}{Ofir Weisse}, \bibinfo{person}{Valeria
  Bertacco}, {and} \bibinfo{person}{Todd Austin}.}
  \bibinfo{year}{2017}\natexlab{}.
\newblock \showarticletitle{Regaining Lost Cycles with HotCalls: A Fast
  Interface for SGX Secure Enclaves}. In \bibinfo{booktitle}{\emph{Proceedings
  of the 44th Annual International Symposium on Computer Architecture}}
  (Toronto, ON, Canada) \emph{(\bibinfo{series}{ISCA '17})}.
  \bibinfo{publisher}{ACM}, \bibinfo{address}{New York, NY, USA},
  \bibinfo{pages}{81--93}.
\newblock
\showISBNx{9781450348928}
\urldef\tempurl%
\url{https://doi.org/10.1145/3079856.3080208}
\showDOI{\tempurl}


\bibitem[Whittaker et~al\mbox{.}(2021)]%
        {whittaker2020scaling}
\bibfield{author}{\bibinfo{person}{Michael Whittaker},
  \bibinfo{person}{Ailidani Ailijiang}, \bibinfo{person}{Aleksey Charapko},
  \bibinfo{person}{Murat Demirbas}, \bibinfo{person}{Neil Giridharan},
  \bibinfo{person}{Joseph~M. Hellerstein}, \bibinfo{person}{Heidi Howard},
  \bibinfo{person}{Ion Stoica}, {and} \bibinfo{person}{Adriana Szekeres}.}
  \bibinfo{year}{2021}\natexlab{}.
\newblock \showarticletitle{Scaling Replicated State Machines with
  Compartmentalization}.
\newblock \bibinfo{journal}{\emph{Proc. VLDB Endow.}} \bibinfo{volume}{14},
  \bibinfo{number}{11} (\bibinfo{date}{jul} \bibinfo{year}{2021}),
  \bibinfo{pages}{2203--2215}.
\newblock
\showISSN{2150-8097}
\urldef\tempurl%
\url{https://doi.org/10.14778/3476249.3476273}
\showDOI{\tempurl}


\bibitem[Xiang et~al\mbox{.}(2021)]%
        {xiang2021strengthened}
\bibfield{author}{\bibinfo{person}{Zhuolun Xiang}, \bibinfo{person}{Dahlia
  Malkhi}, \bibinfo{person}{Kartik Nayak}, {and} \bibinfo{person}{Ling Ren}.}
  \bibinfo{year}{2021}\natexlab{}.
\newblock \showarticletitle{Strengthened Fault Tolerance in Byzantine Fault
  Tolerant Replication}. In \bibinfo{booktitle}{\emph{2021 IEEE 41st
  International Conference on Distributed Computing Systems (ICDCS)}}.
  \bibinfo{publisher}{IEEE}, \bibinfo{address}{DC, USA},
  \bibinfo{pages}{205--215}.
\newblock
\urldef\tempurl%
\url{https://doi.org/10.1109/ICDCS51616.2021.00028}
\showDOI{\tempurl}


\bibitem[Yin et~al\mbox{.}(2003)]%
        {yin2003separating}
\bibfield{author}{\bibinfo{person}{Jian Yin}, \bibinfo{person}{Jean-Philippe
  Martin}, \bibinfo{person}{Arun Venkataramani}, \bibinfo{person}{Lorenzo
  Alvisi}, {and} \bibinfo{person}{Mike Dahlin}.}
  \bibinfo{year}{2003}\natexlab{}.
\newblock \showarticletitle{Separating Agreement from Execution for Byzantine
  Fault Tolerant Services}.
\newblock \bibinfo{journal}{\emph{SIGOPS Oper. Syst. Rev.}}
  \bibinfo{volume}{37}, \bibinfo{number}{5} (\bibinfo{date}{oct}
  \bibinfo{year}{2003}), \bibinfo{pages}{253--267}.
\newblock
\showISSN{0163-5980}
\urldef\tempurl%
\url{https://doi.org/10.1145/1165389.945470}
\showDOI{\tempurl}


\bibitem[Yin et~al\mbox{.}(2019)]%
        {yin2019hotstuff}
\bibfield{author}{\bibinfo{person}{Maofan Yin}, \bibinfo{person}{Dahlia
  Malkhi}, \bibinfo{person}{Michael~K. Reiter}, \bibinfo{person}{Guy~Golan
  Gueta}, {and} \bibinfo{person}{Ittai Abraham}.}
  \bibinfo{year}{2019}\natexlab{}.
\newblock \showarticletitle{HotStuff: BFT Consensus with Linearity and
  Responsiveness}. In \bibinfo{booktitle}{\emph{Proceedings of the 2019 ACM
  Symposium on Principles of Distributed Computing}} (Toronto ON, Canada)
  \emph{(\bibinfo{series}{PODC '19})}. \bibinfo{publisher}{ACM},
  \bibinfo{address}{New York, NY, USA}, \bibinfo{pages}{347--356}.
\newblock
\showISBNx{9781450362177}
\urldef\tempurl%
\url{https://doi.org/10.1145/3293611.3331591}
\showDOI{\tempurl}


\end{thebibliography}
